\documentclass[apj,twocolumn]{emulateapj}

\newcommand{\vo}{V_{0}}

\newcommand{\vio}{(V-Ic)_{0}}
\newcommand{\mm}{(m-M)_{0}}

\newcommand{\feh}{\rm{[Fe/H]}}
\newcommand{\afe}{\rm{[\alpha/Fe]}}

\shorttitle{The stellar populations and structural parameters of UFD galaxies}
\shortauthors{Okamoto et al.}

\submitted{Accepted for pubblication in ApJ}

\begin{document}

\title{The Stellar Populations and Structural Properties of\\
    Ultra Faint Dwarf Galaxies, Canes Venatici I, Bo\"otes I, Canes Venatici II, and Leo IV$^{\dag}$}

\author{
Sakurako Okamoto\altaffilmark{1}, 
Nobuo Arimoto\altaffilmark{2,3}, 
Yoshihiko Yamada\altaffilmark{2}, and 
Masato Onodera\altaffilmark{4}}

\altaffiltext{\dag}{Based on data collected at Subaru Telescope, which is operated by the Nationa Astronomical Observatory of Japan.}

\altaffiltext{1}{Kavli Institute for Astronomy and Astrophysics, Peking University, 100871, China;\email{okamoto@pku.edu.cn}}
\altaffiltext{2}{National Astronomical Observatory of Japan, Osawa 2-21-1, Mitaka, Tokyo, 181-8588, JAPAN}
\altaffiltext{3}{The Graduate University for Advanced Studies, Osawa 2-21-1, Mitaka, Tokyo, 181-8588, JAPAN}
\altaffiltext{4}{Institute	for	Astronomy,	ETH	Z\"urich,	Wolfgang-Pauli-strasse	27,	8093	Z\"urich,	Switzerland}

\begin{abstract}

We take deep images of four ultra faint dwarf (UFD) galaxies, Canes Venatici I (CVn I), Bo\"otes I (Bo\"o I), Canes Venatici II (CVn II), and Leo IV using the Suprime-Cam on the Subaru Telescope.  The colour-magitude diagrams (CMDs) extend below the main sequence turn-offs (MSTOs) and yield measurements of the ages of stellar populations.  The stellar populations of faint three galaxies, Bo\"o I, CVn II and Leo IV dwarf spheroidal galaxies (dSphs) are estimated as old as a Galactic globular cluster M92.  We confirm that Bo\"o I dSph has no intrinsic colour spread in the MSTO, and no spatial difference in the CMD morphology, which indicates that Bo\"o I dSph is composed of an old single stellar population.  One of the brightest UFDs, CVn I dSph, shows a relatively younger age ($\sim$12.6 Gyr) with respect to Bo\"o I, CVn II, and Leo IV dSphs, and the distribution of red horizontal branch (HB) stars is more concentrated toward the centre than that of blue HB stars, suggesting that the galaxy contains complex stellar populations.  Bo\"o I and CVn I dSphs show the elongated and distorted shapes.  CVn II dSph has the smallest tidal radius as a Milky Way satellite and has distorted shape, while Leo IV dSph shows less concentrated spherical shape.   
The simple stellar population of faint UFDs indicates that the gases in their progenitors were removed more effectively than those of brighter dSphs at an occurrence of their initial star formation.  This is reasonable if the progenitors of UFDs belong to the less massive halos than those of brighter dSphs. 

\end{abstract}

\keywords{ Local Group -- galaxies: dwarf -- galaxies: photometry -- galaxies: structure}

\section{Introduction}

The Local Group dwarf spheroidal galaxies (dSphs) in the Milky Way subgroup offer a unique opportunity to investigate galaxy formation and evolution by studying photometric properties of the resolved stellar populations.  Since 2005, more than a dozen new dSphs have been discovered from the Sloan Digital Sky Survey (SDSS) data \citep[e.g.][]{2005ApJ...626L..85W, 2007ApJ...654..897B}.  The surface brightness of these systems are too low to be identified by the photographic plate and therefore are called ultra faint dwarf (UFD) galaxies. 
The UFD galaxies are about 10 to 100 times fainter than the classical dSphs, and are even fainter than most of the Galactic globular clusters.  They have tidally distorted shapes with quite old stellar populations, and yet are the most dark matter (DM) dominated galaxies \citep{2007ApJ...663..948G, 2008A&A...487..103O}.  

Spectroscopic observations have recently revealed that the metallicities of these UFDs are lower than luminous dSphs, and the stellar metalliticy distribution functions (MDFs) are similar to the metal-poor end of the Galactic halo \citep{2008ApJ...685L..43K}.  These results imply that the Galactic halo could have been built from less-luminous satellites similar to these UFDs.  However, because of faint luminosities and apparently large angular sizes, the general features of UFDs, such as the star formation history (SFH) and the detailed structural properties, are still unclear.  Previous studies have shown that the stellar populations of Canes Venatici I (CVn I) dSph, one of the brightest UFD galaxies, can be divided into a kinematically cold, metal-rich component and a hot, metal-poor one \citep{2006MNRAS.373L..70I}, while most of UFDs show a old and metal-poor population alone \citep[e.g.,][]{2006ApJ...647L.111B,2007ApJ...654..897B,2007ApJ...668L..43C}. 

Since the stellar densities of UFDs are extremely low, deep photometry is required to reach below the main sequence turn-off (MSTO) to acquire sufficient number of the members for a statistical analysis of structural properties.  The sensitivity, image quality and wide sky coverage are required to probe both global structure and spatially resolved SFH derived directly from the MSTO.  Therefore, we obtained the images of CVn I, Bo\"otes I (Bo\"o I), Cane Venatici II (CVn II), and Leo IV dSphs by using the Subaru/Suprime-Cam, that is wide enough to cover the entire areas of the dSphs and deep enough to estimate the stellar ages from the old MSTO.  

In this paper, the data reduction are described in Section 2, and the distances of galaxies are estimated in Section 3.  Section 4 presents the colour-magnitude diagrams (CMDs) and discusses the stellar population of each UFD galaxy.  Section 5 shows the spatial distributions and structural properties.  In Section 6, we discuss the population gradients and the age spreads of UFD galaxies, and compare the stellar populations of UFD with those of classical dSphs.  Finally, we summarize our conclusions in Section 7.

\section{Observations and Data Reduction}

\begin{deluxetable*}{lccccccc}
\tablecolumns{7}
\tablewidth{0pt}
\tablecaption{Information about the observations\label{tbl: obs log}}
\tablehead{
\colhead{Galaxy}     & \colhead{Field ID}    &
\colhead{Filter}       & \multicolumn{2}{c}{Short Exposure}    &
\multicolumn{2}{c}{Long Exposure}\\
\cline{4-5} \cline{6-7}
& & & sec$\times$shots & FWHM & sec$\times$shots & FWHM}
\startdata
CVn I & CVN1 & $Ic$ & 30.0s $\times$ 3 & 0\farcs78 & 240.0s $\times$ 13 & 0\farcs82 \\
& &  $V$ & 10.0s $\times$ 3 & 1\farcs00 & 200.0s $\times$ 5 &  1\farcs00 \\
& CVN1\_F$^{*}$ & $Ic$ & 30.0s $\times$ 3 & 1\farcs21 & 240.0s $\times$ 14 & 1\farcs18 \\
& (control field) & $V$ & 10.0s $\times$ 3 & 1\farcs04 & 200.0s $\times$ 5 & 1\farcs12 \\
\tableline\\
CVn II & CVN2 & $Ic$ & 10.0s $\times$ 3 & 0\farcs74 & 200.0s $\times$ 13 & 0\farcs72 \\
& & V & 10.0s $\times$ 3 & 1\farcs56 & 120.0s $\times$ 8 & 1\farcs20 \\
& CVN2\_F$^{*}$ & $Ic$ & 10.0s $\times$ 3 & 1\farcs36 & 200.0s $\times$ 13 &  1\farcs08 \\
& (control field) & $V$ & 10.0s $\times$ 3 & 1\farcs18 & 120.0s $\times$ 5 & 1\farcs18 \\
\tableline \\
Bo\"o I & BOO1 & $Ic$ & 10.0s $\times$ 3 & 0\farcs94  & 200.0s $\times$ 15 & 0\farcs90 \\
& & $V$ & 10.0s $\times$ 3 & 0\farcs98 & 120.0s $\times$ 5 & 1\farcs00 \\
\tableline \\
Leo IV & LEO4 & $Ic$ & 10.0s $\times$ 3 & 1\farcs28 & 200.0s $\times$ 14 & 0\farcs94 \\
& & $V$ & 10.0s $\times$ 3 & 0\farcs92 & 120.0s $\times$ 5 &  0\farcs88 \\
& LEO4\_F$^{*}$ & $Ic$ & 10.0s $\times$ 3 & 1\farcs24 & 200.0s $\times$ 17 & 1\farcs16 \\
& (control field) & $V$ & 10.0s $\times$ 3 & 1\farcs22 & 120.0s $\times$ 5 & 1\farcs24 \\
\enddata
\tablecomments{The control fields for CVn I, CVn II, and Leo IV dSphs are located  at (RA,Dec)$_{\rm{J2000}}$=($13^{h}28^{m}08^{s}.10$, $+33\arcdeg33\arcmin36\farcs49$), ($12^{h}57^{m}14^{s}.66$, $+34\arcdeg19\arcmin30\farcs77$), ($11^{h}33^{m}00^{s}.87$, $-00\arcdeg31\arcmin44\farcs42$), respectively. }
\end{deluxetable*}


\begin{figure}[t]
 \plotone{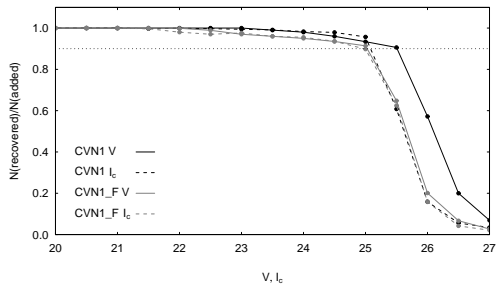}
 \caption{The completeness of the photometry in the CVN1 and CVN1\_F fields as a function of calibrated magnitude.  Solid lines are the completeness derived for $V$-band images and dotted lines for $Ic$-band, and black and gray colour show the CVN1, CVN1\_F fields, respectively.  The horizontal dotted line represents the 90$\%$ completeness limit.}
  \label{fig: completeness of CVn1}
\end{figure}


We took images of CVn I, Bo\"o I, CVn II, and Leo IV dSphs using the Subaru/Suprime-Cam during the nights of 2008 April 2 to April 4 (PI: S. Okamoto; Proposal ID: S08A-022).  The nights of the observing runs were photometric and the seeing ranged from $0.74\arcsec$ to $1.2\arcsec$, except for the short exposures of CVN2 and CVN2\_F fields as shown in Table \ref{tbl: obs log}.  To avoid saturation of bright stars, we took short and long exposure images.  The combination of short and long exposures with Johnson V- and Cousins I-filters allowed us to construct the colour-magnitude diagrams (CMDs) from the bright red giant branch (RGB) to below the old MSTO.  The seeing, exposure time, the airmass of each image, and the ID of each field are listed in Table \ref{tbl: obs log}.  

\begin{figure}[t] 
 \plotone{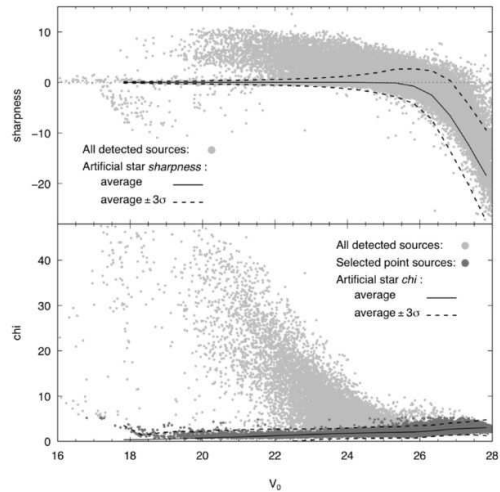}
 \caption{The $chi$ and $sharpness$ of the sources found in $V$-band image of CVN1 field as a function of calibrated magnitude.  $Upper$:  The $sharpness$ of all detected sources are shown in light gray colour.  The solid line shows the mean value of simulated stars of artificial star test, and the dashed lines show the $3\sigma$ deviations.  $Lower$:  The $chi$ of all detected sources and the point-sources selected by $sharpness$ value are shown in light gray and dark gray colour, respectively.  The solid line shows the mean value of simulated star of artificial star test, and the dashed lines show the $3\sigma$ deviations. }
  \label{fig: sharpchi of CVn1 V-band}
\end{figure}

To estimate the contamination of the foreground Galactic stars and the background galaxies, the control field at one degree off from each galaxy centre  at the same Galactic latitude was taken during the same night for CVn I, II, and Leo IV dSphs (see Table \ref{tbl: obs log}).  Unfortunately, a limited time prevented us from getting the images of control field of Bo\"o I dSph.  

\begin{figure*}[t]
 \begin{center}
 \includegraphics[width=500pt]{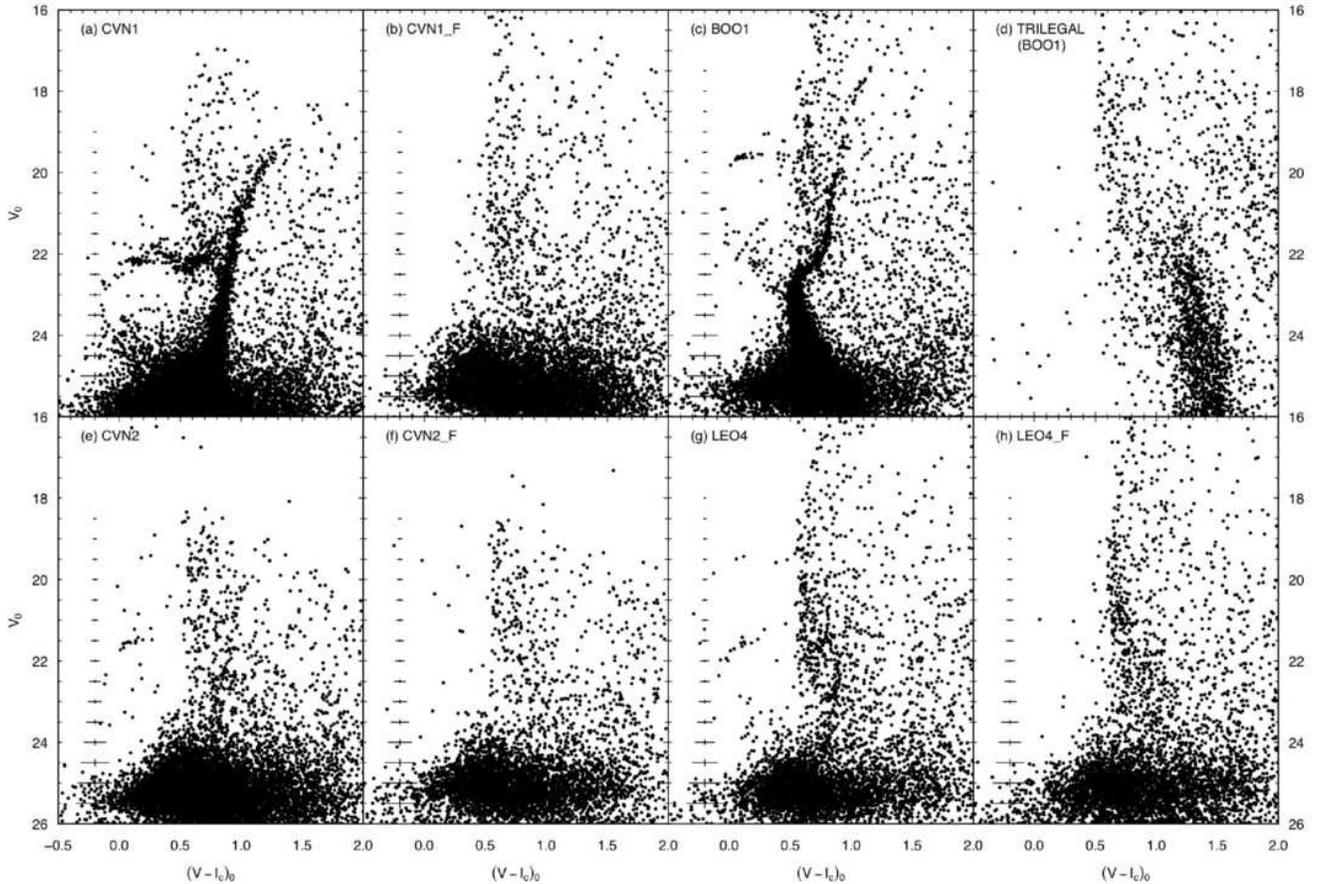}
 \caption{The CMDs of the star-like objects in all observed fields and the simulated CMD. The magnitude errors are estimated by the artificial star test. In Figure \ref{fig: all CMD}d, TRILEGAL model is used to simulate the CMD of Galactic foreground stars in the direction of BOO1 field.}
  \label{fig: all CMD}
  \end{center}
\end{figure*}

The raw data were processed using a pipeline software SDFRED dedicated to the Suprime-Cam \citep{2002AJ....123...66Y, 2004ApJ...611..660O} in usual manner.  Each raw image was bias-subtracted and trimmed, flat-fielded by self-flat image, corrected for distortion and atmospheric dispersion, checked and matched for the point-spread-function (PSF), sky-subtracted, and combined.  For the processed images, the DAOPHOT in IRAF package was used to obtain the PSF photometry of the resolved stars \citep{1987PASP...99..191S}.  

SCAMP and SExtractor were used to compute astrometric solutions for the processed images with astrometric standard stars selected from the SDSS catalogues \citep{1996A&AS..117..393B, 2006ASPC..351..112B}.  The instrumental magnitudes of sources in the images were calibrated to the standard Johnson-Cousins photometric system using the photometric standard stars of Landolt \citep{1992AJ....104..340L}, observed during each run.  The average extinction in the direction of each field is taken from \citet{1998ApJ...500..525S}.  The assumed extinction law is R$_{V}$=3.1 \citep{1989ApJ...345..245C} and A$_I$/A$_V$=0.594 \citep{1998ApJ...500..525S}.

To estimate the accuracy and incompleteness of the photometric catalogues, the artificial star tests were performed on the images of each field with the ADDSTAR routine in DAOPHOT.  The percentage of detected point sources at a given magnitude has been calculated by adding artificial stars, which were made from the PSF model, to the images, and the resulting images are processed in the same way as for the original ones.  The 7000 artificial stars are added to each image for every 0.5 magnitude interval from 18 mag to 27 mag.  This number is smaller than one-tenth of the detected sources of each image to avoid the blending of artificial stars with real sources.  The detection ratio of the test, N(recovered)/N(added) of $V$- and $Ic$-band images of CVn I dSph is plotted in Figure \ref{fig: completeness of CVn1}.  The artificial star test shows that our photometry is at least 90$\%$ complete at 25 mag in both $V$- and $Ic$-band in the whole region of all galaxies.

The mean photometric errors are based on the difference between the input magnitude and the output magnitude of the simulated stars in the artificial star test.  These errors are plotted in the CMD of each field in the following sections.  

To separate stars from the extended sources and noise-like objects, we used the image sharpness statistic $sharpness$ and the goodness of fit statistic $chi$ parameters of DAOPHOT, both are efficient to select point sources by the artificial star test.  In the upper panels of Figure \ref{fig: sharpchi of CVn1 V-band}, the $sharpness$ of all detected sources in the $V$-band long-exposure image of CVN1 field are shown.  We select the sources, of which $sharpness$ are within $3\sigma$ of the mean of the simulated stars (dashed lines).   The $chi$ values of all sources and the selected point sources are shown in the lower panels of Figure \ref{fig: sharpchi of CVn1 V-band}.  The $chi$ values of most of the selected sources are within $3\sigma$ of the mean of the simulated stars.  The star/galaxy separation degrades at magnitude fainter than $V\sim 24.5$ and $Ic\sim 24.0$, on average.  

Figures \ref{fig: all CMD} show the resulting CMDs of the star-like objects found in all central and control fields of galaxies. The CMD of the central field of each galaxy is extended below the MSTO (Figures \ref{fig: all CMD}a,c,e,g). In Figure \ref{fig: all CMD}a, both red and blue HB of CVn I dSph appear with the well defined RGB and the blue straggler (BS) candidates which are not found in that of the control field (Figure \ref{fig: all CMD}b).  The blue HB, the tight MS, and the BS candidates of Bo\"o I dSph are seen in Figure \ref{fig: all CMD}c.  The central CMD of CVn II dSph (Figures \ref{fig: all CMD}e) has similar distribution to that of the control fields (Figures \ref{fig: all CMD}f), but the blue HB and the weak signal of RGB can be found in Figures \ref{fig: all CMD}e.  The CMDs of Leo IV dSphs (Figures \ref{fig: all CMD}g,h) have the similar properties to those of CVn II dSph, but the BS candidates are also seen in Figure \ref{fig: all CMD}g.  The lack of bright stars in Figures \ref{fig: all CMD}a,e,f are due to the poor observing condition.  We discuss the stellar populations in detail in Section \ref{sec: pop}. 

In Figure \ref{fig: all CMD}d, TRILEGAL\footnote{http://trilegal.ster.kuleuven.be/cgi-bin/trilegal} model code \citep{2005A&A...436..895G} is used to simulate the CMD of Galactic foreground stars in the direction of BOO1 field.  We use this model CMD to estimate the contamination of Bo\"o I dSph in Section \ref{sec: structure}.  In the model CMD, the halo MS stars are distributed at $\vio > 0.6$ mag, which are also seen in those of the observed fields. The faint red objects distributed at the $\vo > 24.5$ in the observed CMDs are the unresolved background galaxies, which does not appear in the model CMD.  The background galaxies at $\vio > 0.1$ in Figure \ref{fig: all CMD}b are brighter than those found in the CVN1 field, probably due to the poor observing condition.

\section{Distances}

\subsection{Canes Venatici I dSph}
The magnitudes of the horizontal blanch (HB) and the RGB tip (TRGB) were used to the estimate the distances of the galaxies.  For CVn I dSph, 36 blue HB (BHB) stars in the range of $22.05 < \vo < 22.35 $ and $0.1 < \vio < 0.3$ are used to derive the average magnitudes $V_{\rm{BHB}}=22.17 \pm 0.05$ and the red HB (RHB) derived from 113 stars with $22.05 < \vo < 22.35$ and $0.58 < \vio < 0.85$ as $V_{\rm{RHB}}=22.13 \pm 0.06$.  The mean value is $V_{\rm{HB}} = 22.15 \pm 0.06$, which is in good agreement with the average magnitude of the RR Lyrae stars $V_{\rm{RR}} = 22.17 \pm 0.02$ estimated by \citet{2008ApJ...674L..81K}.  The calibration of absolute magnitude $M_{V,\rm{RR}}$ is a function of the metallicity $\feh$ \citep{2003LNP...635..105C}.  Using spectroscopic metallicity $\feh = -2.08 \pm 0.02$ derived by \citet{2008ApJ...685L..43K}, the absolute magnitude of RR Lyrae stars is estimated to be $M_{V,\rm{RR}} = 0.47 \pm 0.05$, from which the distance modulus of CVn I dSph is determined to $\mm = 21.68 \pm 0.08$.  

The $Ic$-band magnitude of the TRGB can also be used as a distance indicator.  $M_{Ic,\rm{TRGB}}$ is not sensitive to the metallicity at $\feh < -0.7$, nor to the age if it is older than several Gyrs \citep[e.g.][]{1993ApJ...417..553L}.  The number of the bright RGB stars in a UFD, however, is usually too small and the bright RGB is heavily buried in the foreground contamination.  But thanks to the relatively bright luminosity, the location of the TRGB of CVn I dSph is estimated.  The position of TRGB is found at $Ic_{\rm{TRGB}} = 17.91 \pm 0.06$ from bright three RGB stars with $17.9 < Ic_{0} < 18.0$ and $1.35 < (V-Ic)_{0} < 1.47$.  The distance modulus of CVn I dSph is then derived from the calibration of $M_{Ic,\rm{TRGB}}$ as a function of the metallicity $\rm{[M/H]}$ \citep{1993ApJ...414..580S, 2004A&A...424..199B}.  Using the spectroscopic metallicity $\feh = -2.08 \pm 0.02$ and assuming the $\afe \sim 0$ as found in classical dSphs \citep[e.g. ][]{2003AJ....125..684S}, the absolute magnitude of TRGB in Ic-band is estimated as $M_{Ic,\rm{TRGB}} = -3.51$, from which the distance modulus is derived as $(m-M)_{0} = 21.48 \pm 0.17$.  We assume the same $\afe$ as those of Draco and Ursa Minor dSphs, because the luminosity of CVn I dSph is similar to these faint classical dSphs, although spectroscopic confirmation is required.  If we adopt $\afe$ of Galactic globular clusters ($\afe \sim +0.3$), the distance modulus of CVn I dSph becomes $(m-M)_{0} = 21.90 \pm 0.17$.  These two values are slightly smaller 
($(m-M)_{0,\afe\sim0.0} = 21.48 \pm 0.17$) and larger ($(m-M)_{0, \afe \sim +0.3}=21.90 \pm 0.17$) than the value estimated based on the HB magnitude ($(m-M)_{0, \rm{HB}}=21.68 \pm 0.08$).  It is due to the uncertainties of the $\afe$ and $Ic_{\rm{TRGB}}$.  Therefore, we adopt the distance modulus of $\mm = 21.68 \pm 0.08$ ($216 \pm 8$ kpc) estimated by the HB luminosity.  Our estimate is consistent with the previous estimate, $\mm = 21.69 \pm 0.10$, by \citet{2008ApJ...672L..13M}.

\subsection{Bo\"otes I dSph}
The mean $V$-band magnitude of the BHB stars in Bo\"o I dSph is derived from 17 stars with $19.5 < \vo < 19.7 $ and $0.0 < \vio < 0.3$ as $V_{\rm{BHB}}=19.63 \pm 0.04$.  The average RR Lyrae star magnitude is extrapolated as $V_{\rm{RR}} \sim 19.43 \pm 0.09$, by using the magnitude difference of BHB and RR Lyrae stars ($\Delta V=0.2 \pm 0.08$) derived from metal-poor ([Fe/H]$<-2.0$) and old ($>12$ Gyr) isochones \citep{2008A&A...482..883M}.  With the spectroscopic average metallicity [Fe/H] = $-2.51 \pm 0.13$ derived by \citet{2008ApJ...689L.113N}, the absolute magnitude of RR Lyrae is $M_{V,\rm{RR}} = 0.36 \pm 0.07$; the distance modulus of Bo\"o I dSph is determined as $\mm = 19.07 \pm 0.11$ ($65 \pm 3$ kpc), which is in good agreement with $\mm = 18.96 \pm 0.12$ ($62 \pm 4$ kpc) given by \citet{2008ApJ...674L..81K}.

\subsection{Canes Venatici II dSph}
In the case of CVn II dSph, it is hard to identify each evolutional stage, in particular the HB from the CMD in Figure \ref{fig: pop CMD of CVN2}.  Therefore, the CMD of the central region was used to derive the magnitude of the HB.  The mean $V$-band magnitude of the BHB stars is derived from 8 stars with $21.3 < \vo < 21.9 $ and $0.0 < \vio < 0.3$, within $5\arcmin$ (corresponding to three times of half-light radius $r_{h}$, estimated by \citealt{2007ApJ...654..897B}) from the centre of CVn II dSph, as $V_{\rm{BHB}}=21.64 \pm 0.06$.  The average RR Lyrae star magnitude is extrapolated as  $V_{\rm{RR}} \sim 21.44 \pm 0.09$ in the same manner as the case of Bo\"o I dSph.  Considering the spectroscopic average metallicity, [Fe/H] = $-2.19 \pm 0.05$, derived by \citet{2008ApJ...685L..43K}, the absolute magnitude of RR Lyrae is $M_{V,\rm{RR}} = 0.43 \pm 0.07$.  The distance modulus of CVn II dSph is determined as $\mm = 21.01 \pm 0.11$ ($159 \pm 8$ kpc), which is in good agreement with previous estimation, $\mm = 21.02 \pm 0.06$ ($160\pm5$ kpc), based on the RR Lyrae magnitude \citep{2008ApJ...675L..73G}. 

\subsection{Leo IV dSph}
Leo IV dSph has similar luminosity and distance to CVn II dSph.  Therefore, the distance of Leo IV dSph is estimated in the same manner as CVn II dSph.  The mean $V$-band magnitude of the BHB stars is $V_{\rm{BHB}}=21.53 \pm 0.06$ which is derived from 7 stars with $21.0 < \vo < 21.7 $ and $0.0 < \vio < 0.3$, within 5\arcmin (corresponding to twice of $r_{h}$, estimated by \citealt{2007ApJ...654..897B}) from the centre of Leo IV dSph.  The average RR Lyrae star magnitude is extrapolated as $V_{\rm{RR}} = 21.33 \pm 0.09$.  With the spectroscopic average metallicity, [Fe/H] = $-2.58 \pm 0.08$, derived by \citet{2008ApJ...685L..43K}, the absolute magnitude of RR Lyrae is $M_{\rm{V},\rm{RR}} = 0.34 \pm 0.09$; therefore the distance modulus of Leo IV dSph is $\mm = 20.99 \pm 0.12$ ($158 \pm 8$ kpc), which is in excellent agreement with the previous estimation, $\mm = 20.94 \pm 0.07$ ($154 \pm 5$ kpc), based on the magnitude of RR Lyrae  \citep{2009ApJ...699L.125M}.

\section{Stellar Populations} \label{sec: pop}

\subsection{Canes Venatici I dSph}
CVn I dSph is one of the brightest UFDs \citep{2006ApJ...643L.103Z}.  The HB morphology and RGB slope in the CMD look like those of classical dSphs such as Draco dSph.  Figure \ref{fig: CMD of CVN1} shows the de-reddened CMDs of the star-like objects found in the central and control fields of CVn I dSph.  In Figure \ref{fig: CMD of CVN1}a, the well defined RGB appears with the tip at $\vo \sim 19.2$ and $\vio \sim 1.3$, and the HB stars are seen at $\vo \sim 22.2$.  Both BHB and RHB found at $\vio \sim 0.1$ and $\vio \sim 0.7$, respectively.  The MS stars can be traced below $\vo \sim 25$, and significant number of blue straggler (BS) candidates are found at $\vio < 0.3$ and $\vo \sim 24$.  On the other hand, the foreground stars are heavily distributed at $\vio > 0.6$ mag, and the background galaxies are found as the distribution at the $\vo > 24.5$.  These objects are also found in the CMDs of all other target and control fields.  The faint red objects at $\vio > 0.1$ in the control field are brighter than those found in the CVN1 field, probably due to the poor observing condition. 

\begin{figure*}
 \begin{center}
 \includegraphics[width=400pt]{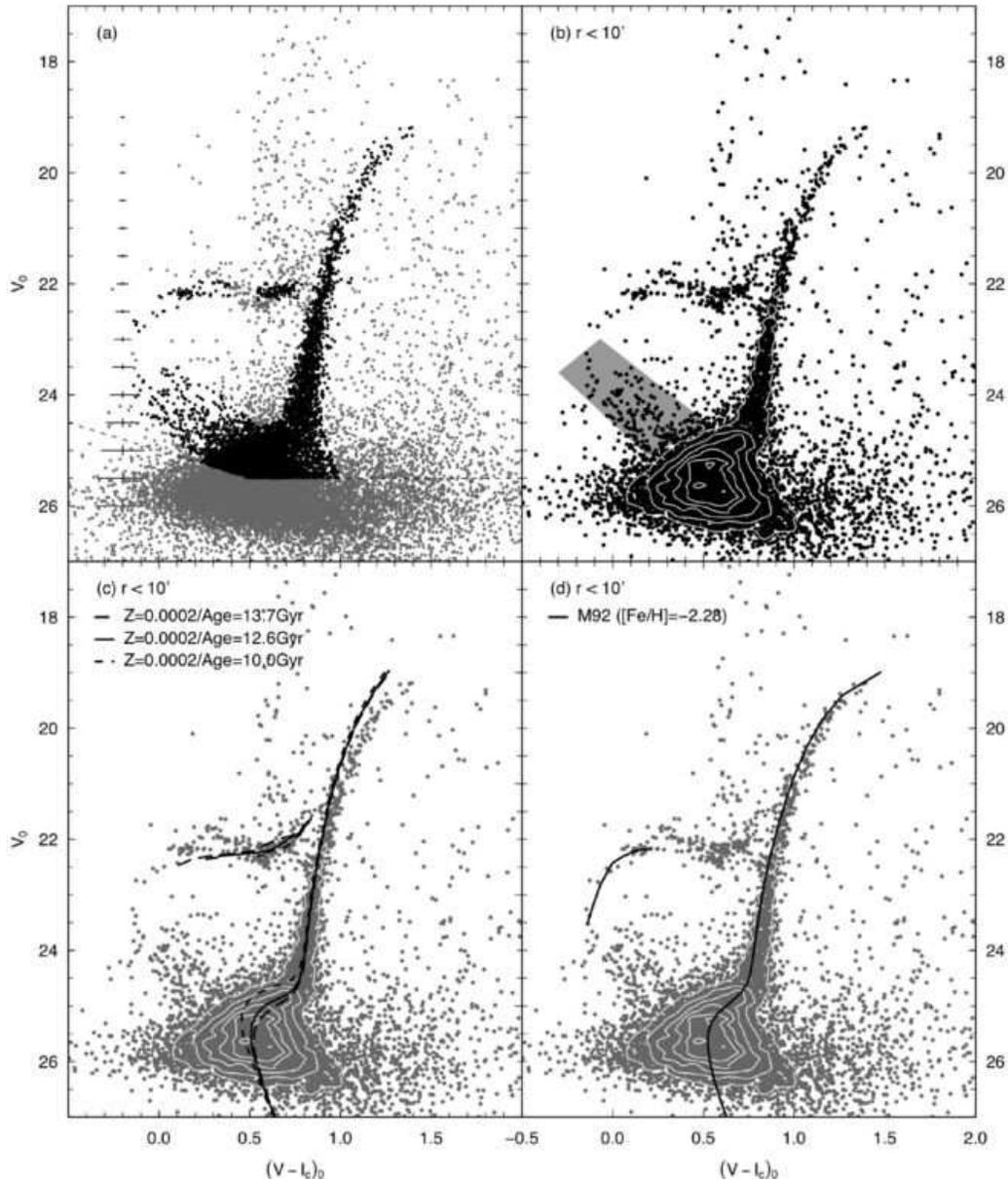}
 \caption{The $\vio$-$\vo$ CMDs of whole region (a) and within $r_{h}$ region (b-d) of CVn I dSph.  In Figure \ref{fig: pop CMD of CVN1}a, the dashed line presents the 90\% detection limit of photometry.  Black points are the member candidates of CVn I dSph which are used to estimate the structural properties.  On Figure \ref{fig: pop CMD of CVN1}c and \ref{fig: pop CMD of CVN1}d, the theoretical isochrones and the fiducial sequence of Galactic globular cluster M92 are overlaid. }
  \label{fig: pop CMD of CVN1}
  \end{center}
\end{figure*}

Figure \ref{fig: pop CMD of CVN1} shows the CMDs of whole region (Figure \ref{fig: pop CMD of CVN1}a) and the central region within $r_{h}$ ($< 10$ \arcmin, Figures \ref{fig: pop CMD of CVN1}b-\ref{fig: pop CMD of CVN1}d) of CVn I dSph.  In Figure \ref{fig: pop CMD of CVN1}a, the gray colour shows all star-like objects, and the black points present the member candidates brighter than the 90 $\%$ detection limits of $V$- and $Ic$-band photometry (dashed line).  We use these candidates for studying the spatial distributions of CVn I dSph in Section \ref{sec: structure}.  In Figures \ref{fig: pop CMD of CVN1}b-\ref{fig: pop CMD of CVN1}d, contours are over-plotted for the purpose of clarity.  In Figures \ref{fig: pop CMD of CVN1}c and \ref{fig: pop CMD of CVN1}d, theoretical isochrones and the fiducial sequence of the metal-poor Galactic globular cluster M92 are overlaid.  The fiducial sequence of M92 was taken from \citet{1998AJ....115..693J} and \citet{2008AJ....135..682C}.  The metallicity of M92 is $\feh = -2.28$ \citep{1996AJ....112.1487H}, and the age is estimated as 14.2$\pm$1.2 Gyr \citep{2007AJ....133.2787P}.  
The fiducial sequence of M92 from \citet{2008AJ....135..682C} was converted from $g', r', i'$ to Johnson-Cousins $V,Ic$ system using the transformation given by \citet{2006A&A...460..339J}.  We adopt the distance modulus $\mm = 14.67$ and the reddening corrections $E(B-V) = 0.02$ for M92 \citep{1996AJ....112.1487H}.  The locations of MSTO and RGB on the CMD are similar to those of M92, suggesting that the average metallicity of CVn I dSph is $\feh\simeq -2.3$, which is consistent with the spectroscopic estimate by \citet{2008ApJ...685L..43K}. 

By overlaying theoretical isochrones in Figure \ref{fig: pop CMD of CVN1}c, the average age of stellar population was estimated.  Padova isochrones of $Z=0.0002$ and 10.0, 12.6, 13.7 Gyr \citep{2008A&A...482..883M} are used by shifting to the distance of CVn I dSph.  The metallicity we adopt corresponds to $\feh = -2.2$ with $\afe = +0.3$, and $\feh = -2.0$ with $\afe =0.0$.  Figure \ref{fig: pop CMD of CVN1}c shows that the fiducial sequence from MSTO to RGB of CVn I dSph is best reproduced by the isochrone of 12.6 Gyr.  

\citet{2008ApJ...672L..13M} found a dichotomy in the stellar populations of CVn I dSph which shows an old ($> 10$Gyr), metal-poor ($\feh \sim -2.0$) and spatially extended population along with a younger ($\sim 1.4-2.0$ Gyr), metal-rich and centrally concentrated one.  They suggested that the blue plume of stars at $B-V \sim 0.1$ and $23.5 < V < 25.0$ indicated the presence of a possible young stellar population.  The deep Suprime-Cam photometry, however, shows no evidence for young stars, instead it shows BSs clearly (see Section \ref{subsec: BS}).  

The HB morphology of CVn I dSph, which shows both BHB and RHB, is similar to that of the brighter classical dSphs such as Fornax and Sculptor dSphs.  \citet{2004ApJ...617L.119T} and \citet{2006A&A...459..423B,2008ApJ...681L..13B} showed that Fornax and Sculptor dSphs have the two distinct HB populations; the spatially extended metal-poor BHB stars and the centrally concentrated metal-rich RHB stars, which also appear to be kinematically distinct.  The stellar population structure of CVn I dSph will be discussed in more detail in Section \ref{subsec: pop radial of CVn1}.

\begin{figure*}
 \begin{center}
 \includegraphics[width=400pt]{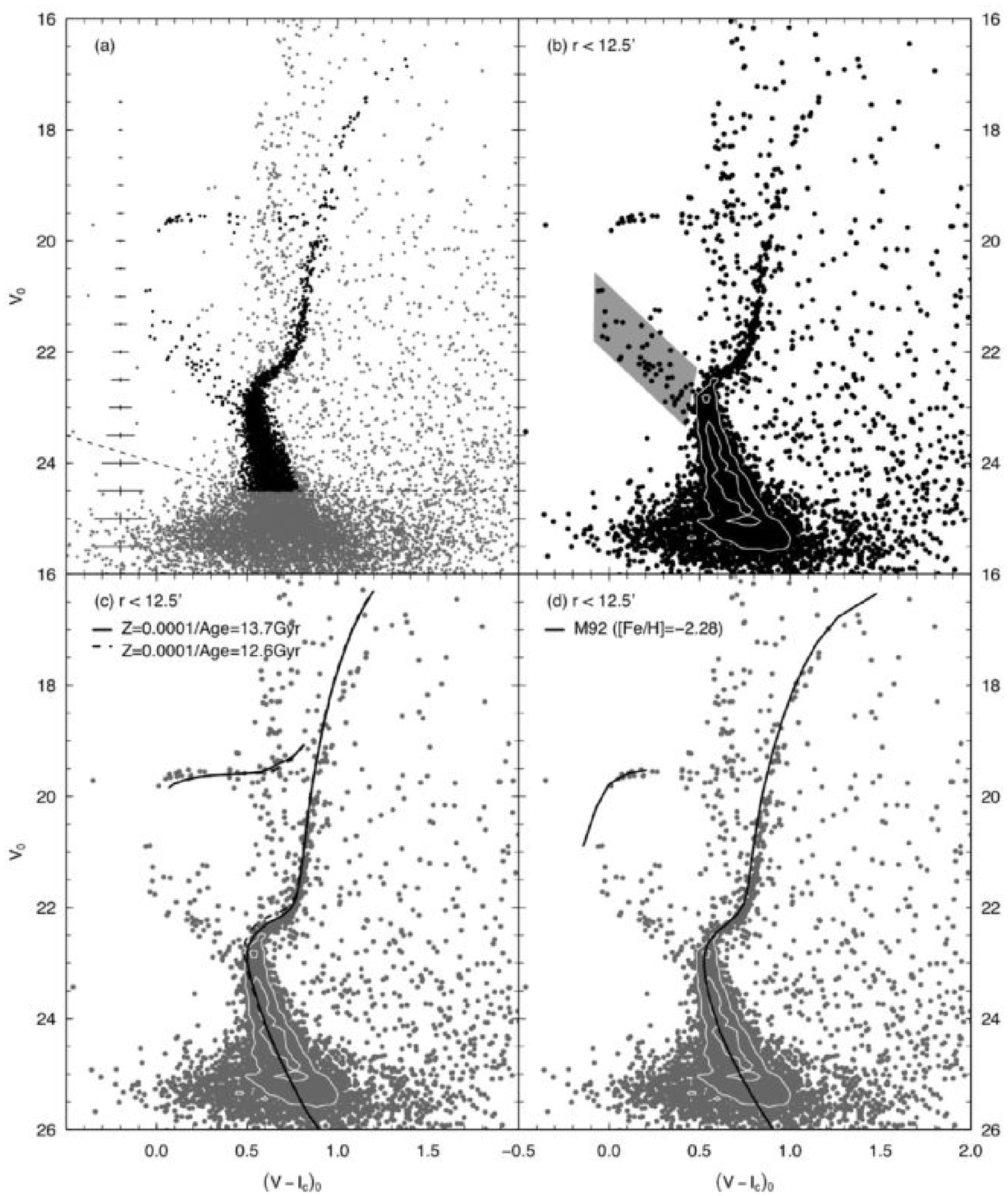}
 \caption{Same as Figure \ref{fig: pop CMD of CVN1}, but for Bo\"o I dSph. }
  \label{fig: pop CMD of BOO1}
  \end{center}
\end{figure*}

\subsection{Bo\"otes I dSph}

Bo\"o I dSph has a moderate luminosity ($M_{V} \sim -5.8$) of UFD.  Thanks to its small distance, the tight MS is clearly seen in the CMD in Figure \ref{fig: pop CMD of BOO1}.  Figure \ref{fig: pop CMD of BOO1} shows the CMD of whole observed region (Figure \ref{fig: pop CMD of BOO1}a), and within 12.5\arcmin~(corresponding to $r_{h}$) region (Figures \ref{fig: pop CMD of BOO1}b-\ref{fig: pop CMD of BOO1}d),  with the theoretical isochrones (Figure \ref{fig: pop CMD of BOO1}c) and the fiducial sequence of M92 (Figure \ref{fig: pop CMD of BOO1}d) in the same manner as the case of CVn I dSph.  The sequences of MS, RGB and HB of Bo\"o I dSph are quite similar to those of M92, suggesting that the average metallicity of Bo\"o I dSph is $\feh \simeq -2.3$, which is consistent with the spectroscopic estimate by \citet{2008ApJ...689L.113N}.  

In Figure \ref{fig: pop CMD of BOO1}c, Padova isochrones with Z=0.0001 and 12.6 and 13.7 Gyr \citep{2008A&A...482..883M} are shifted to the distance of Bo\"o I dSph.  This metallicity corresponds to $\feh = -2.5$ with $\afe = +0.3$, and $\feh = -2.3$ with $\afe =0.0$.  Figure \ref{fig: pop CMD of BOO1}c shows that the fiducial sequence of MS is best reproduced by the isochrone of 13.7 Gyr, consistent with the age roughly estimated from the comparison with M92 in Figure \ref{fig: pop CMD of BOO1}d.  In fact, even the oldest isochrone shows slightly bluer colour at MSTO than that of Bo\"o I dSph.  This difference would be reduced if we adopt the more metal-rich isochrone, but it is not likely that the spectroscopically confirmed metallicity is underestimated significantly \citep{2008ApJ...689L.113N}.  Consequently, the main population of Bo\"o I dSph is estimated to be older than that of CVn I dSph.  Figure \ref{fig: pop CMD of BOO1}a indicates that the MSTO colour width is quite narrow, which implies no age spread of stars in Bo\"o I dSph.  We discuss it in Section \ref{subsec: age of Boo1}.

\subsection{Canes Venatici II dSph}

\begin{figure}[t]
\plotone{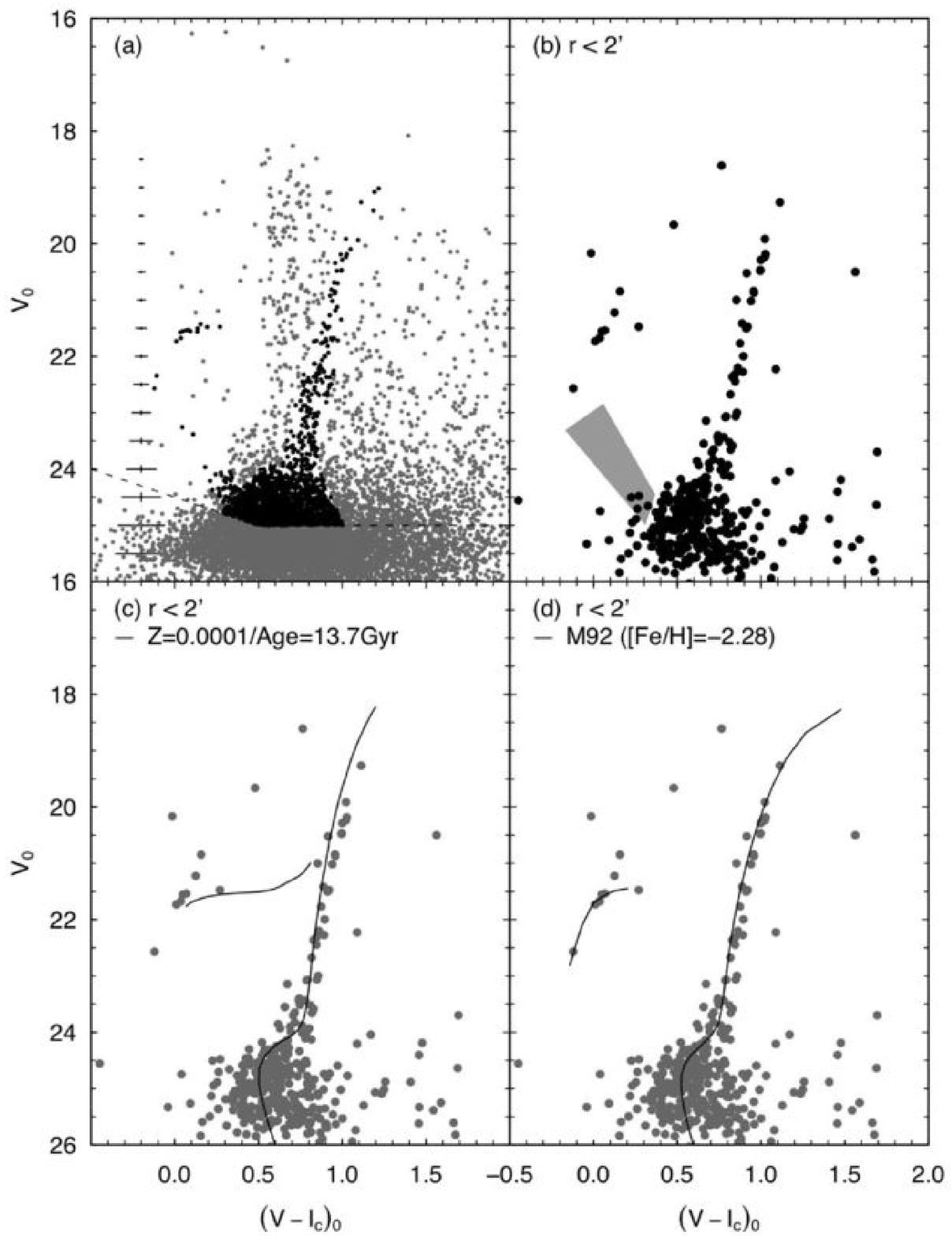}
 \caption{Same as Figure \ref{fig: pop CMD of CVN1}, but for CVn II dSph.}
  \label{fig: pop CMD of CVN2}
\end{figure}

CVn II dSph is relatively faint and compact.  The existence of BHB stars implies that CVn II dSph has an old and metal-poor stellar population.  The number of member candidates of CVn II dSph is extremely small, so that we use the central region within $r_{h}$ to estimate the stellar population.  Figure \ref{fig: pop CMD of CVN2} shows the CMDs of CVn II dSph in the same manner as that of CVn I dSph.
In the central CMD (Figures \ref{fig: pop CMD of CVN2}b-d), the stellar sequences of MS, RGB and BHB are clearly seen.  The narrow RGB is seen at $0.6 < \vio < 1.0$ and $19 < \vo < 24$, but the bright RGB stars is not seen, because of the lack of short exposure images of CVN2 field.  The BHB stars are identified at $\vo \sim 21.5$ mag and $\vio \sim 0.25$.  The MS stars can be found at $\vo < 24.0$ and $\vio \sim 0.6$ mag.  Interestingly, there are no bright BS candidates in the central CMD (Figures \ref{fig: pop CMD of CVN2}b-d), either due to a real paucity of BS stars or to a poor statistics.  

Figure \ref{fig: pop CMD of CVN2}d shows that the fiducial line of MSTO to RGB of CVn II dSph is similar to those of M92, suggesting that CVn II dSph is quite old and metal poor ([Fe/H] $\simeq -2.3$).  The latter is consistent with the spectroscopic estimate by \citet{2008ApJ...685L..43K}.  The average age of stars found in the central ($< r_{h}$) region is estimated as 13.7 Gyr by overlaying Padova isochrones in Figure \ref{fig: pop CMD of CVN2}c.  This age is the same as that of Bo\"o I dSph.  Therefore, most of stars in CVn II dSph are estimated to have formed at the same epoch when Bo\"o I dSph was born.

\begin{figure}[t]
\plotone{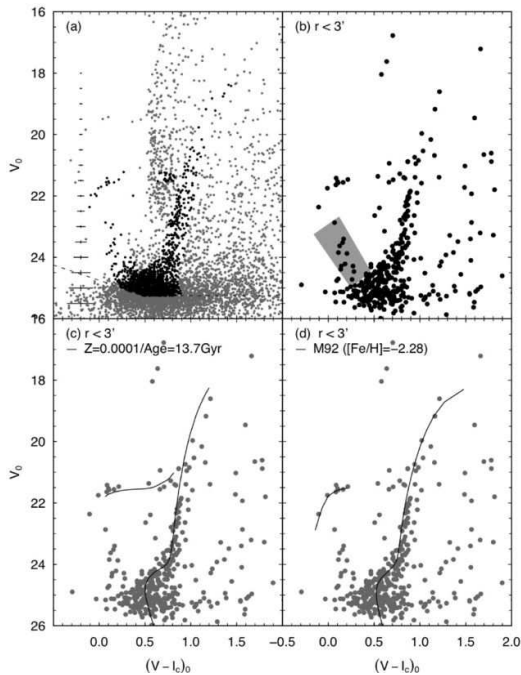}
 \caption{Same as Figure \ref{fig: pop CMD of CVN1}, but for Leo IV dSph.}
  \label{fig: pop CMD of LEO4}
\end{figure}

\subsection{Leo IV dSph}

Figure \ref{fig: pop CMD of LEO4} shows the CMDs of Leo IV dSph in the same manner as CVn I dSph. The stellar distribution in the CMD is similar to that of CVn II dSph, and thus it is rather difficult to identify stellar sequences in Figure \ref{fig: pop CMD of LEO4}a.  In the CMD of the central region (Figure \ref{fig: pop CMD of LEO4}b-d), MS, RGB, BHB and BS stars are clearly seen.  The narrow RGB is seen at $0.6 < \vio < 1.2$ and $18.5 < \vo < 24$.  The BHB stars are seen at $\vo \sim 21.5$ and $\vio \sim 0.25$.  The MS stars can be found at $\vo < 24.0$ and $\vio \sim 0.6$, and the BS candidates at $ 23.0 < \vo < 24.5$ and $0.0 <\vio < 0.4$.  The major difference between Leo IV (Figure \ref{fig: pop CMD of LEO4}b) and CVn II (Figure \ref{fig: pop CMD of CVN2}b)  dSphs is the existence of BS and RHB candidates in Leo IV dSph.  These populations do not appear in the central CMD of CVn II dSph.  The spatial distribution of RHB candidates is not concentrated but uniformly distributed throughout the observed area; they could be the foreground stars.  

The ridge line from MS to RGB of Leo IV dSph is similar to that of M92, suggesting that the average metallicity of Leo IV dSph is [Fe/H] $\simeq -2.3$, which is consistent with the spectroscopic estimate \citep{2008ApJ...685L..43K}.  Similar to Bo\"o I and CVn II dSphs, the average age of stellar population of Leo IV dSph is estimated as 13.7 Gyr by overlaying Padova isochrones in Figure \ref{fig: pop CMD of LEO4}c.  

\citet{2010ApJ...718..530S} recently have pointed out that blue plume stars in Leo IV dSph are young ($\sim 2$Gyr) populations and are equivalent to those found in CVn I dSph by \citet{2008ApJ...672L..13M}.  The spatial distribution of these stars, however,  clearly indicates that they are the BS stars as we will discuss in Section \ref{subsec: BS}.

\section{Structural Properties} \label{sec: structure}


\begin{figure*}
 \begin{center}
 \includegraphics[width=350pt]{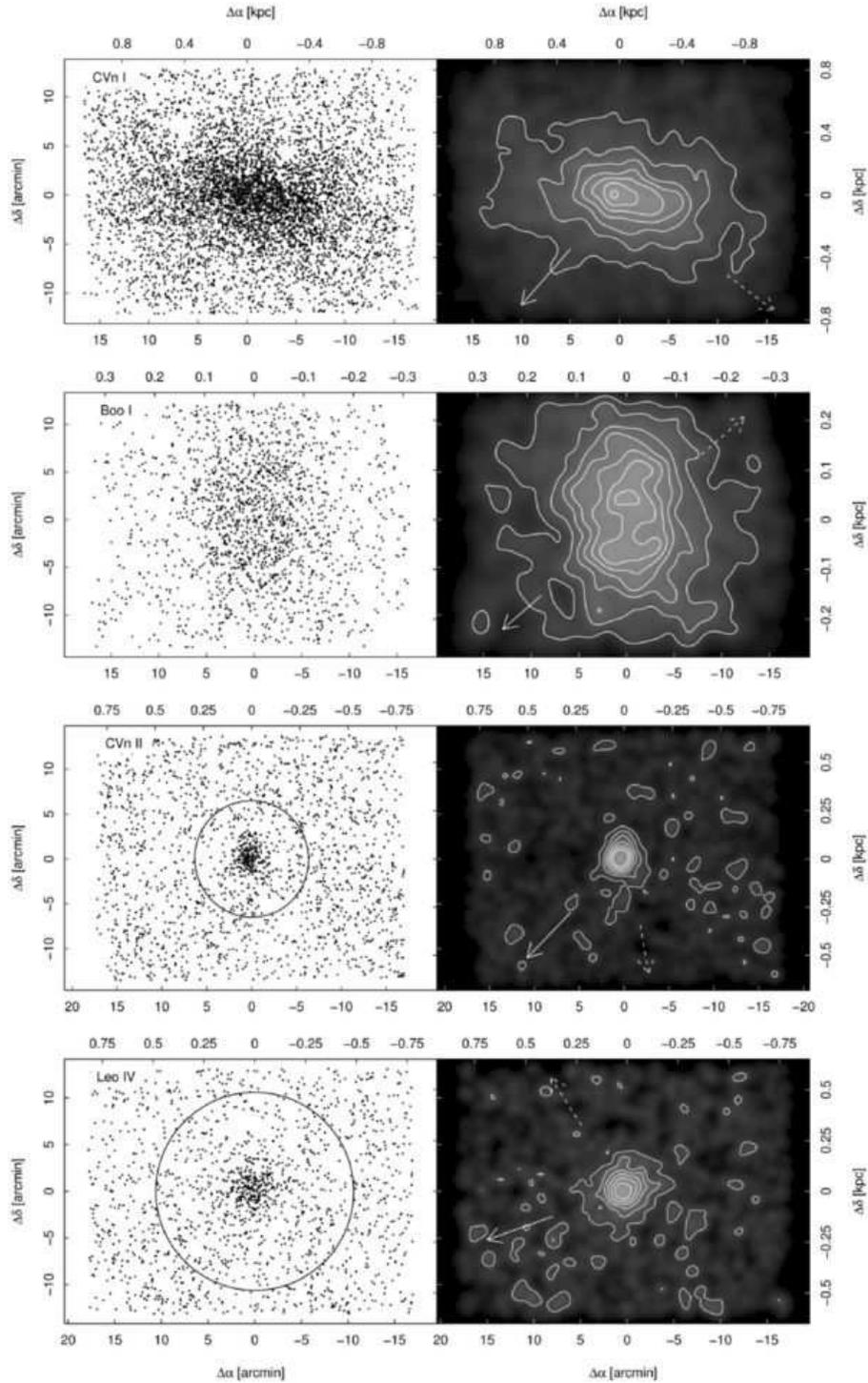}
 \caption{The spatial distribution of CVn I (upper), Bo\"o I (upper-middle), CVn II (lower-middle) and Leo IV (lower) dSphs.  $Left$: The spatial distribution of the member candidates selected from each CMD.  {\it The tidal radii of CVn II and Leo IV dSphs are shown as solid lines. }$Right$: The isodensity contour of the member candidates. The solid and dashed arrows indicate the directions of Galactic centre (Sgr A) and Galactic latitude toward the disk respectively. }
  \label{fig: spatial map}
  \end{center}
\end{figure*}


Figure \ref{fig: spatial map} give the spatial distributions and the density contour maps of UFDs, which show various morphologies.  CVn I and Bo\"o I dSphs show elongated and distorted shapes.  CVn II dSph has relatively high concentration, while Leo IV dSph shows smooth and less concentrated spherical shape.  The plotted sources are selected as the candidate stars of MS, RGB, HB and BS from the CMDs, shown as black points in Figure \ref{fig: pop CMD of CVN1}a, \ref{fig: pop CMD of BOO1}a, \ref{fig: pop CMD of CVN2}a and \ref{fig: pop CMD of LEO4}a.  These stars are binned and smoothed by the Gaussian kernels to make the contour map.  The contour levels are 2, 4, 6, 8, 12, and 14$\sigma$ above the background level.  The maps cover the central 2 kpc $\times$ 1.5 kpc of CVn I dSph, 600 pc $\times$ 500 pc of Bo\"o I dSph, and  1.7 $\times$ 1.3 kpc of CVn II and Leo IV dSphs, respectively.  

Because we do not have the control field of Bo\"o I dSph, we choose the selection limit brighter than the degradation level of our star/galaxy separation to eliminate the background galaxy contamination in Bo\"o I dSph.  Therefore, the spatial distribution of Bo\"o I dSph is constructed by these member candidates, which are bright enough to exclude the background contamination found in the distribution at $\vo > 24.5$.  The foreground contamination is estimated by using the model CMD constructed by TRILEGAL shown in Figure \ref{fig: all CMD}. 

\begin{figure*}[th] 
 \begin{center}
 \includegraphics[width=150pt,angle=-90,clip]{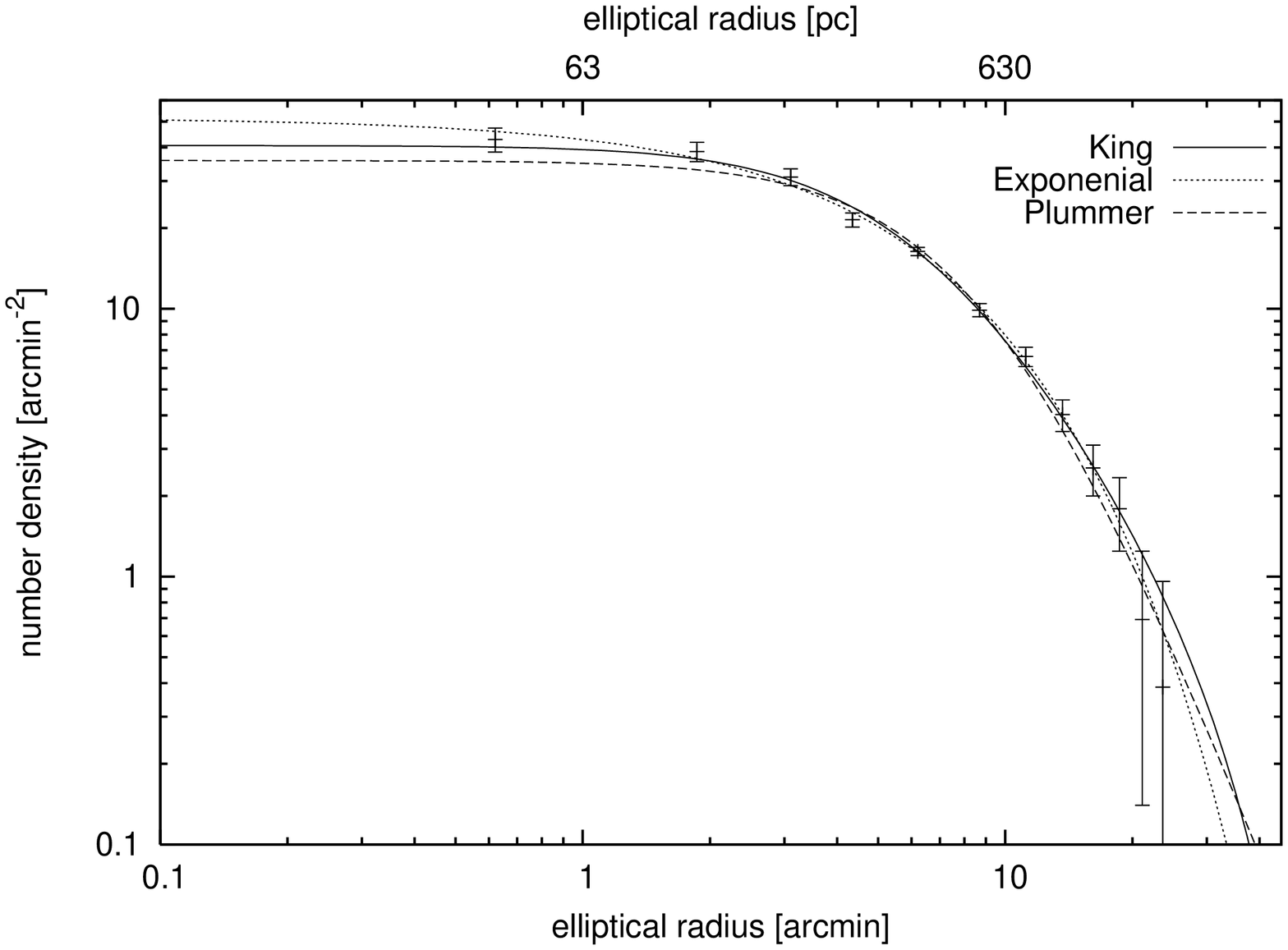}
 \includegraphics[width=150pt,angle=-90,clip]{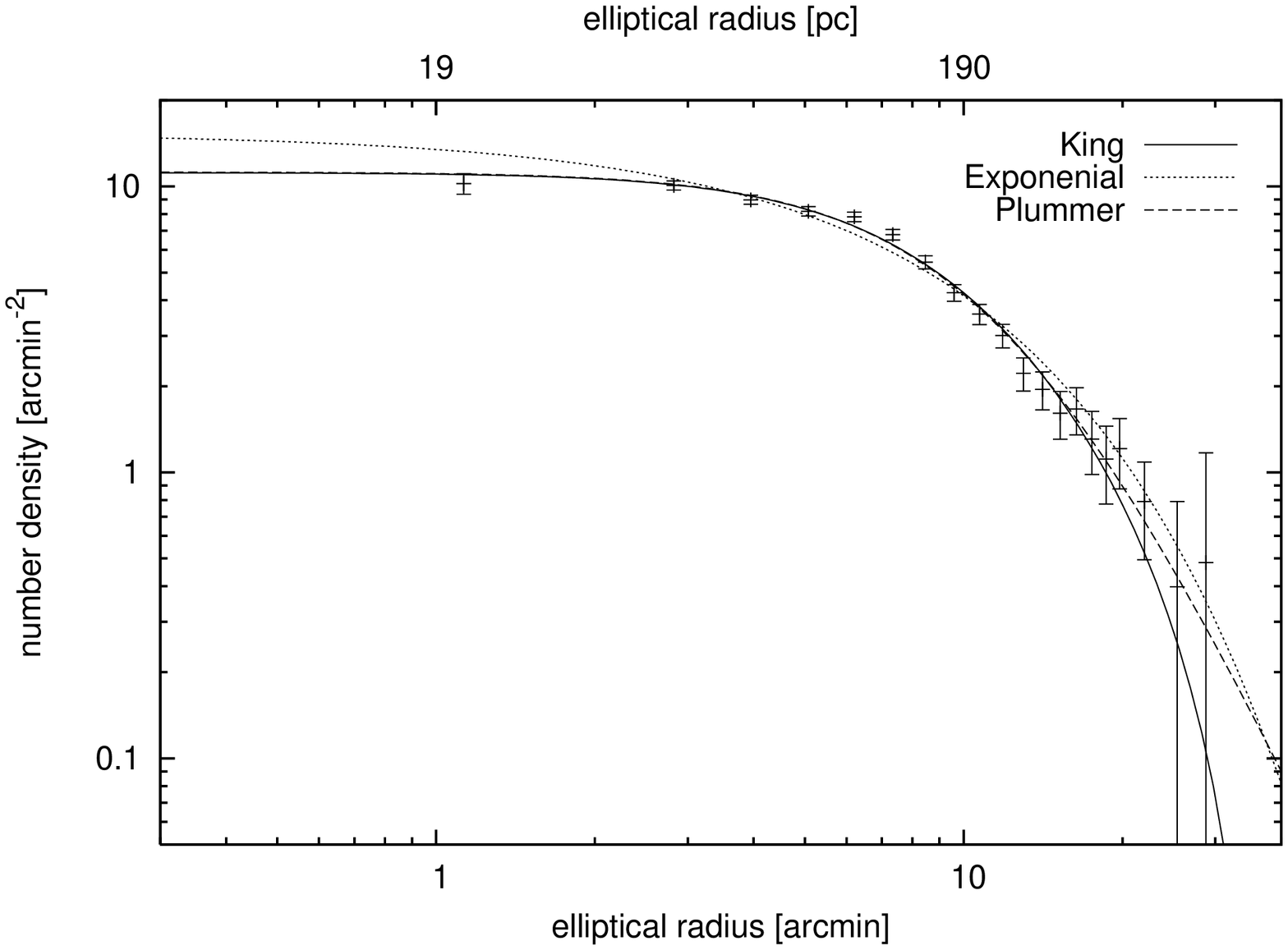} \\
 \vspace{5pt}
 \includegraphics[width=150pt,angle=-90,clip]{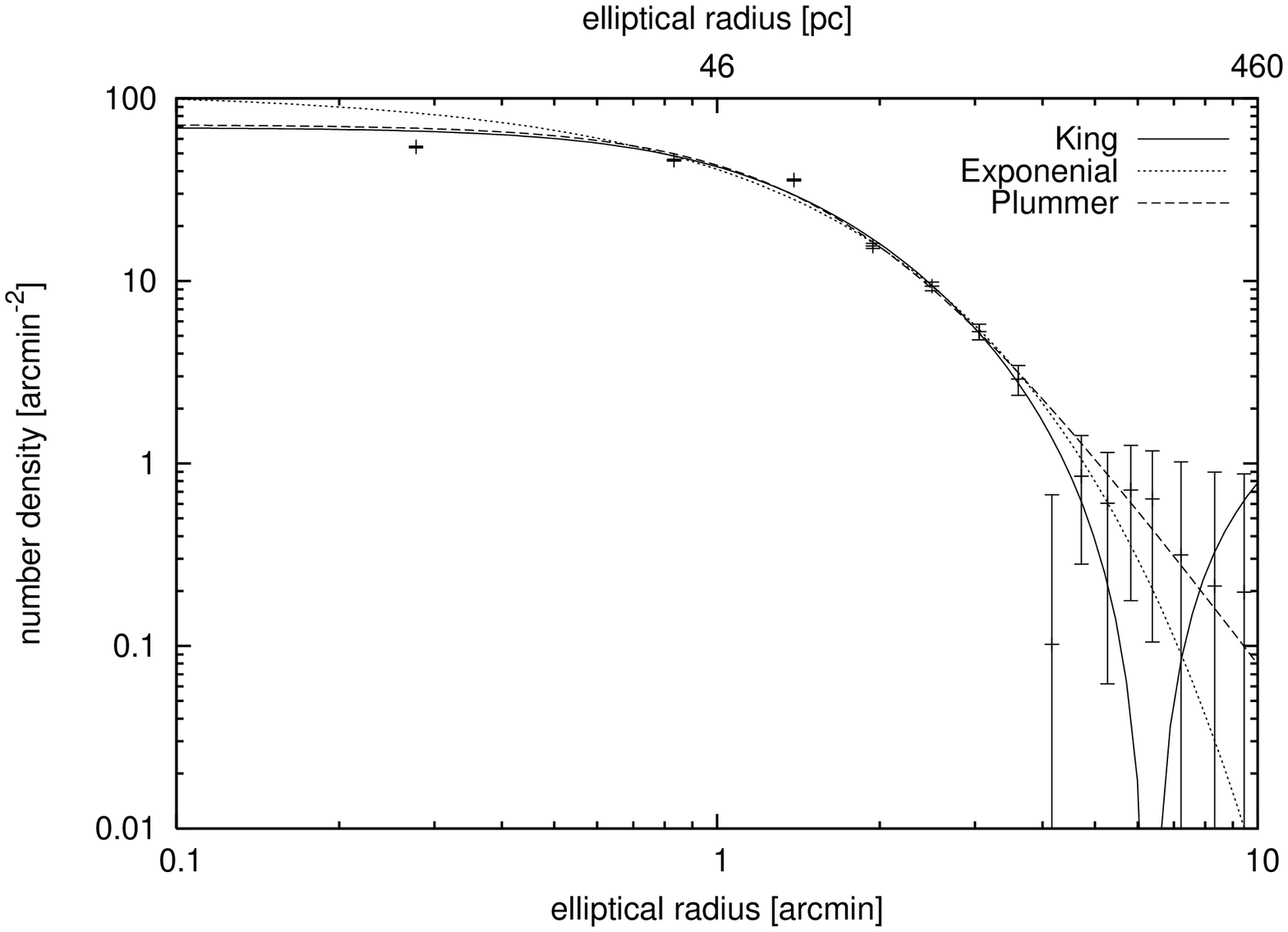}
 \includegraphics[width=150pt,angle=-90,clip]{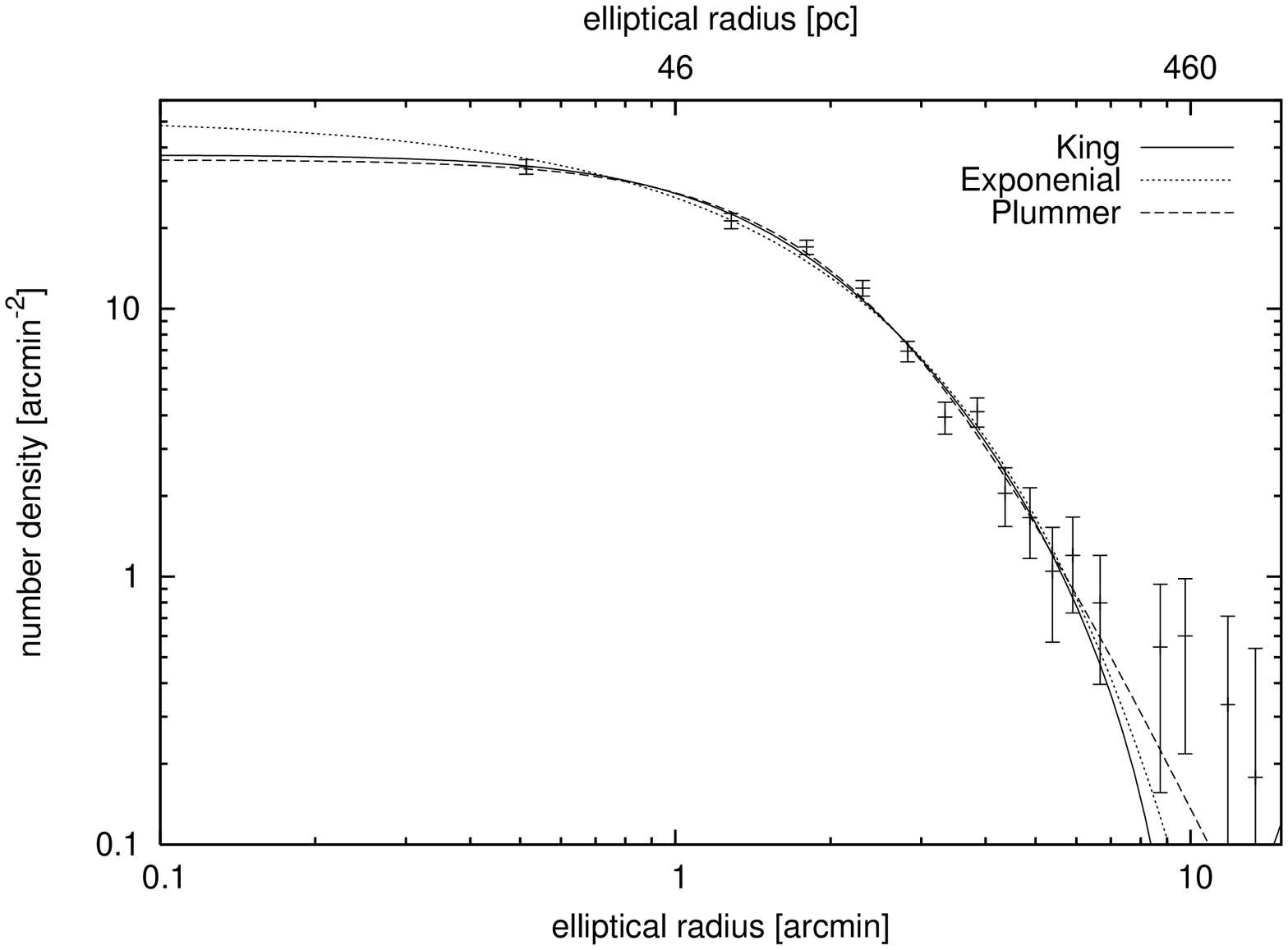} \\ 
 \caption{The radial profiles of UFD galaxies derived by calculating the average contamination-corrected number density of CMD-selected stars within elliptical annuli. The profiles of CVn I, Bo\"o I, CVn II, and Leo IV dSphs are shown in the upper-left, upper-right, lower-left, and lower-right panels, respectively.  The number densities of foreground/background objects, 2.3, 0.1, 3.2 and 2.45~arcmin$^{-2}$ for CVn I, Bo\"o I, CVn II, and Leo IV dSph, respectively, are subtracted.  The standard King, exponential and Plummer profiles are fitted as the solid, dotted, and dashed line, respectively. }
  \label{fig: radial profile}
  \end{center}
\end{figure*}

The member candidates of each galaxy are used to derive the centroid from the density-weighted first moment of the spatial distribution, and the average ellipticity and the position angle using the three density-weighted second moments \citep[e.g.,][]{1980JBIS...33..323S}.  Figure \ref{fig: radial profile} shows the radial profiles derived from the average number density within elliptical annuli after correcting the effect of the contamination.  The number density of foreground/background objects in the direction of each galaxy is estimated by counting the number of objects within the same criterion in the control field CMD, or the model CMD constructed by TRILEGAL in case of Bo\"o I dSph.  We fit the radial profile with the standard King model \citep{1962AJ.....67..471K},

\begin{equation}
	\Sigma (r)=\Sigma_{\rm{K},0} \Biggl( \frac{1}{\sqrt{1+(r/r_{c})^{2}}} - \frac{1}{\sqrt{1+(r_{t}/r_{c})^{2}}} \Biggr)^{2},
	\label{eq:king}
\end{equation}

\noindent with the exponential model \citep{1968adga.book.....S},

\begin{equation}
	\Sigma (r)=\Sigma_{\rm{E},0} \exp \biggl\{ - \biggl( \frac{r}{r_e} \biggr) \biggl\},
	\label{eq:exp}
\end{equation}

\noindent and with the Plummer model,

\begin{equation}
	\Sigma (r)=\Sigma_{\rm{P},0} \frac{b^2}{(b^2+r^2)^2},
	\label{eq:plummer}
\end{equation}

\noindent to estimate the core radius $r_{c}$, the tidal radius $r_{t}$, and the half light radius $r_{h}$ ($=1.68 \times r_{e} = b$).  Hereafter we use the elliptical radius

\begin{equation}
	r = \Bigl( x^{2} + \frac{y^{2}}{(1-e)^{2}} \Bigr)^{1/2},
	\label{eq:elliptical radius}
\end{equation}

\noindent where $e$ is the ellipticity of the galaxy, $x$ and $y$ are the coordinates aligned with the major and minor axes, respectively.  The estimated $r_{h}$ agrees well with the previous estimates derived from the SDSS data \citep{2008ApJ...684.1075M}.  The core and tidal radii of these galaxies except for CVn II dSph are estimated by this study for the first time.  The tidal radius of CVn I dSph, $r_{t}\sim 3.5$~kpc, is similar to that of Sextans dSph ($r_{t}=4.0$ kpc) which is the largest among the Galactic satellites \citep{1998ARA&A..36..435M}, while the tidal radius of CVn II dSph, $r_{t}\sim300$ pc, is the smallest.  

The absolute magnitude $M_{V}$ is calculated by using the luminosity of member candidates selected from the CMDs within $r_{h}$ in each galaxy.  First, the luminosity is corrected for contamination, that is, the luminosity from the foreground/background objects estimated by the control field CMDs is subtracted.  Next, the luminosity from fainter MS stars is added, and then the resulting luminosity is doubled to estimate the total flux.  The errors include uncertainties of these corrections and the error of $r_{h}$ estimates.  The best-fitting structural parameters are listed in Table \ref{tbl:structure}. 

The contour maps of UFDs show various appearances.  CVn II dSph has an asymmetric shape and extended structure toward south.  The core of CVn II dSph looks spherical shape, but the outer region is elongated toward north and south, which indicate on-going tidal disruption.  Leo IV dSph, on the other hand, shows the less concentrated than CVn II dSph and the spherical shape.  The slight overdensities are found in the southwest of Leo IV dSph.  In CVn I dSph, the peak of stellar density is somewhat offset toward west from the expected centre.  The observed regions do not cover the whole area but cover only the area within the half light radius of CVn I and Bo\"o I dSphs, so it is unclear whether these features really reflect genuine extent of these galaxies or not.  Quite irregular shapes of these galaxies, however, indicate that they are suffering from strong tidal effects from the Milky Way.  The shapes of UFDs apparently have no correlation with the projected direction of Galactic centre and the Galactic latitude which are shown as solid and dashed arrows, respectively, in Figure \ref{fig: spatial map}. 

The tidal disturbance in the stellar distributions of UFDs provides a clue to the understanding the properties of dark matter and the mass profiles of satellite galaxies.  There is no apparent substructure or tidal debris around CVn II and Leo IV dSphs, the observed area of which are extending beyond the tidal radii (Figure \ref{fig: spatial map}).  However, the stellar densities show slight excesses at the edge of these galaxies in Figure \ref{fig: radial profile}.  Similar excesses were also found in the outer regions of several other Galactic satellites, and theoretical studies show that  Galactic tides acting on cuspy halo profiles of satellites and/or the CDM model tend to make higher central densities and excesses in outer regions of satellites in comparison with the cases of cored halos and/or the warm dark matter model \citep[e.g.][]{2002MNRAS.336..119M, 2010MNRAS.406.1290P}.  Although spectroscopic confirmation is required to reveal whether the slight excesses in CVn II and Leo IV dSphs are real, the stellar density profiles of UFDs could provide some constraints on the properties of dark matter in UFDs. 

\begin{deluxetable*}{lcccc}
\tablecolumns{5}
\tablewidth{0pt}
\tablecaption{Propeties of UFD galaxies \label{tbl:structure}}
\tablehead{
\colhead{Parameter} & \colhead{CVn I} & \colhead{Bo\"o I} & \colhead{CVn II} & \colhead{Leo IV}}
\startdata
RA (J2000) & $13^{h}28^{m}01\fs4$ & $14^{h}00^{m}05\fs4$ & $12^{h}57^{m}08\fs5$ & $11^{h}32^{m}56\fs0$ \\
Dec (J2000) & $+33\arcdeg33\arcmin07\farcs7$ & $+14\arcdeg30\arcmin02\farcs0$ & $+34\arcdeg19\arcmin17\farcs4$ & $-0\arcdeg32\arcmin24\farcs7$ \\
Position angle & $78\fdg6$ & $14\fdg2$ & 9\fdg45 & 36$^{\circ}$.9\\
Ellipticity & 0.30 & 0.22 & 0.23 & 0.04\\
$r_{c}$ & 5\farcm62 $\pm$ 0\farcm40 & 10\farcm3 $\pm$ 0\farcm9 & 1\farcm50 $\pm$ 0\farcm16 & 1\farcm79 $\pm$ 0\farcm17 \\
$r_{t}$ & 54\farcm5 $\pm$ 12\farcm9 & 37\farcm4 $\pm$ 5\farcm5 & 6\farcm33 $\pm$ 0\farcm53 & 10\farcm6 $\pm$ 1\farcm08 \\
$r_{h}$ (exponential) & 8\farcm99 $\pm$ 0\farcm20 & 12\farcm8 $\pm$ 0\farcm7 & 1\farcm77 $\pm$ 0\farcm10 & 2\farcm44 $\pm$ 0\farcm10 \\
$r_{h}$ (Plummer) & 9\farcm23 $\pm$ 0\farcm39 & 12\farcm5 $\pm$ 0\farcm3 & 1\farcm85 $\pm$ 0\farcm09 & 2\farcm55 $\pm$ 0\farcm8 \\
$M_{V}$ & $-7.93$ $\pm$ 0.2 & $-5.92 \pm 0.2$ & $-5.37 \pm 0.2$ & $-4.97 \pm 0.2$ \\
$\mm$ & 21.68 $\pm$ 0.08 & 19.07 $\pm$ 0.11 & 21.01 $\pm$ 0.11 & 20.99 $\pm$ 0.12 \\
Distance [kpc] & 216 $\pm$ 8 & 65 $\pm$ 3 & 159 $\pm$ 8 & 158 $\pm$ 8 \\
\enddata
\end{deluxetable*}

\section{Discussions}

It is well known that the classical dSphs show various star formation histories \citep[e.g.,][]{2003AJ....126..218M,2006A&A...459..423B,2009ARA&A..47..371T}.  Population complexities are found in bright dSphs such as Fornax, Leo I, and Carina dSphs, which contain intermediate age ($<$ a few Gyr) and old ($> 10$ Gyr) stars.  Faint classical dSphs such as Ursa Minor show essentially single old population \citep[e.g.,][]{2002AJ....123.3199C}.  In the case of UFDs, we have shown that Bo\"o I, CVn II, and Leo IV dSphs have genuine old population, while the brighter UFD, CVn I dSph has relatively younger population.  This variety of stellar population is probably related to the internal and/or external mechanisms that regulate the star formation of very faint galaxies in the past.

\subsection{The Population Gradient in CVn I dSph} \label{subsec: pop radial of CVn1}

\begin{figure}[t] 
\begin{center}
\includegraphics[width=150pt,angle=-90,clip]{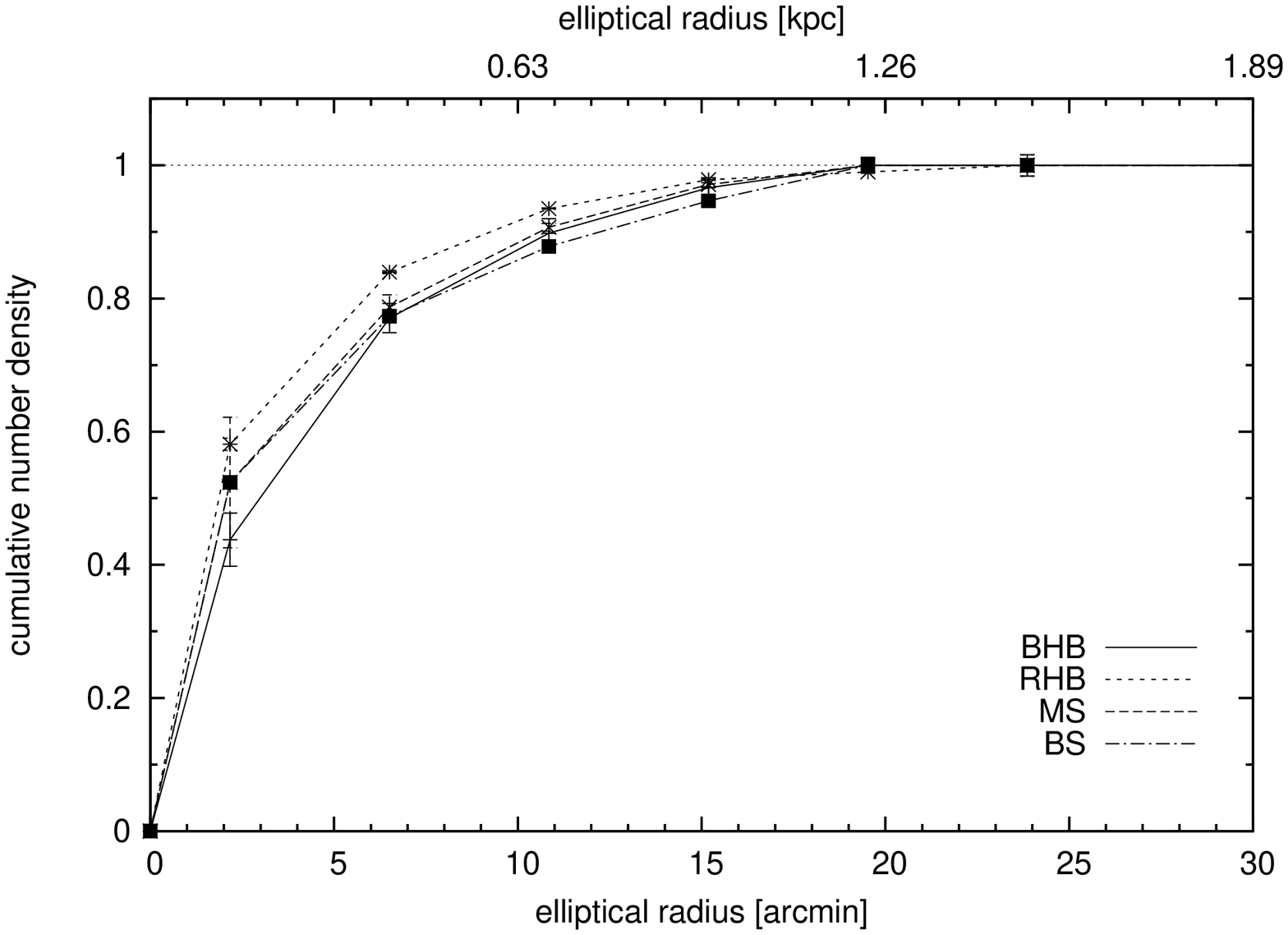}
 \caption{The cumulative radial profiles of each evolutionary phase, BHB, RHB, MS and BS stars in CVn I dSph derived by calculating the average number density within elliptical annuli, are shown in solid, dashed, long dashed, dotted, and dotted-dashed lines, respectively. }
  \label{fig: radial pop of CVn I}
  \end{center}
\end{figure}

With the structural parameters estimated above, the spatial distributions of stellar components of CVn I dSph are derived.  Figure \ref{fig: radial pop of CVn I} presents the cumulative radial distributions of BHB, RHB, MS and BS stars of CVn I dSph.  The foreground/background contaminations are estimated from the control CMD and corrected.  In Figure \ref{fig: radial pop of CVn I}, the BHBs are clearly more extended than other components, and the RHBs are more concentrated toward the galaxy centre.  The colour of HB star reflects the age and metallicity; the metal-rich or younger HB stars become redder than the metal-poor or older HB stars.  Therefore, this radial difference of HB morphology suggests the population gradient in CVn I dSph.  The spatial distributions of HBs are consistent with the result of \citet{2006MNRAS.373L..70I} and \citet{2007MNRAS.380..281M}, who revealed the presence of two kinematically distinct populations in CVn I dSph.  

From these results, we conclude that CVn I dSph has the ancient ($> 10$ Gyr) but at least two populations of different metallicity, spatial distribution, and kinematics. 

\subsection{Age Spread in Bo\"o I dSph ?} \label{subsec: age of Boo1}


\begin{figure}[t] 
 \plotone{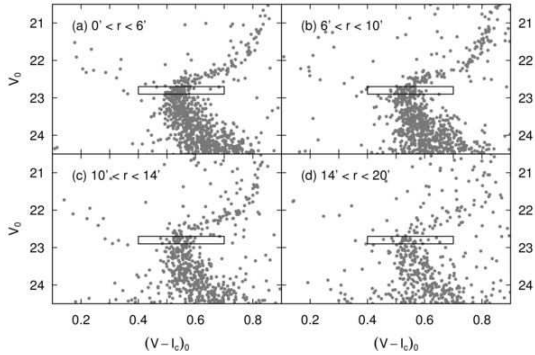}
 \caption{The CMD of central to outer regions of Bo\"o I dSph.  Each CMD includes the stars within the region of elliptical radius of a) $0\arcmin<$r$<6\arcmin$, b) $6\arcmin<$r$<10\arcmin$, c) $10\arcmin<$r$<14\arcmin$, d) $14\arcmin<$r$<20\arcmin$. The rectangle region shows the magnitude and colour ranges to derive the MSTO colour distributions of Figures \ref{fig: bright MS histogram of 4region of Boo1}.}
  \label{fig: CMDs of 4region of Boo1}
\end{figure} 



\begin{figure}[t] 
 \plotone{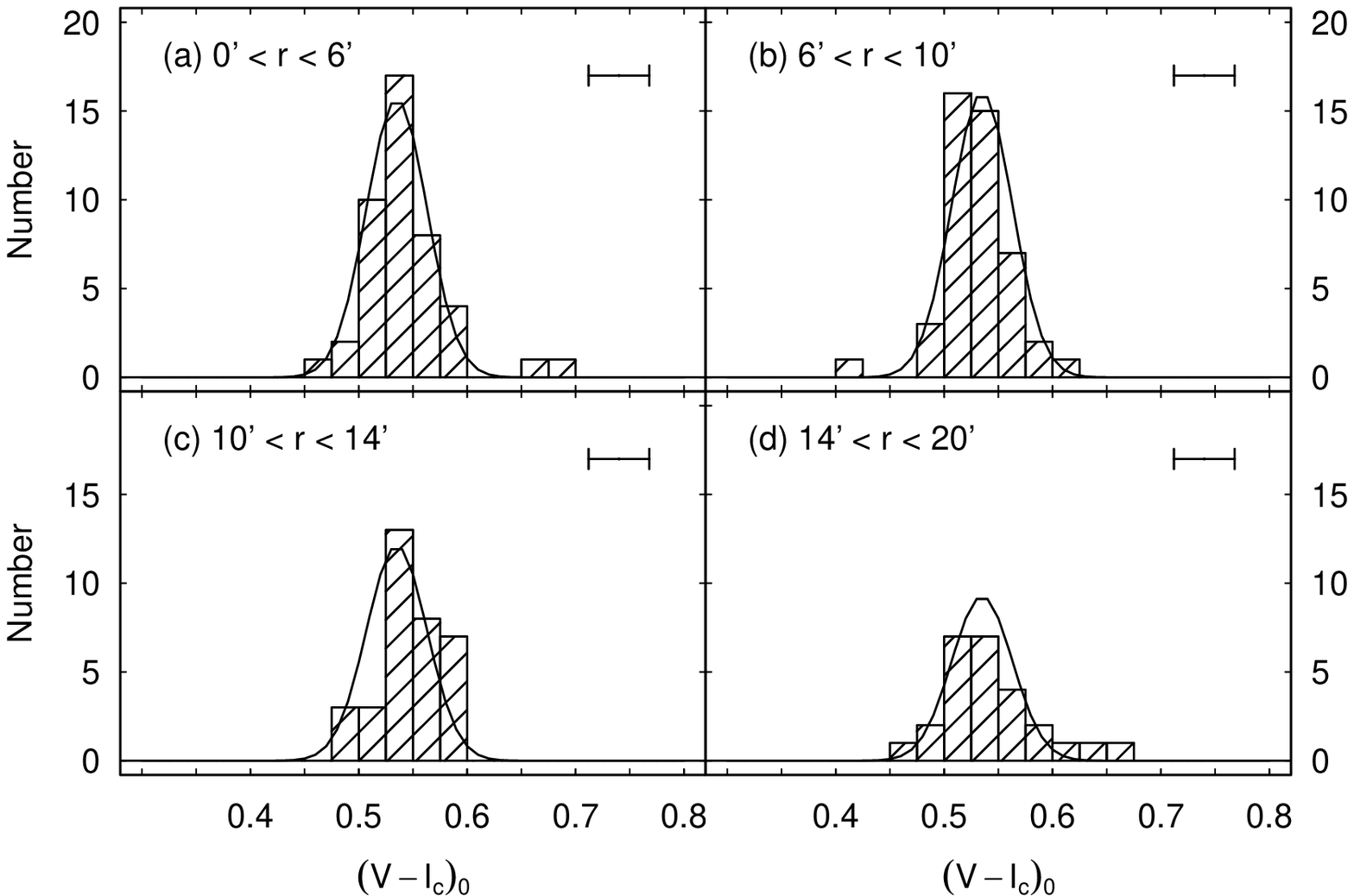}
 \caption{The colour histogram of MSTO stars in Bo\"o I dSph within the black rectangle indicating in each CMD of Figure \ref{fig: CMDs of 4region of Boo1}, with the colour bins of 0.025mag.  The solid lines are Gaussians with $\sigma=0.03$ equal to the photometric errors. }
  \label{fig: bright MS histogram of 4region of Boo1}
\end{figure}


The deep Suprime-Cam photometry presents the sequence of MS to RGB in the CMD of Bo\"o I dSph.  The width of well-defined RGB to MS is quite narrow, and the CMD morphology agrees well with the old metal-poor M92.  This result implies that Bo\"o I dSph has purely old single population as Galactic old globular clusters.  With the limited resolution of our photometry and the coarse grid of theoretical isochrones, it is difficult to conclude whether this galaxy had formed before the reionization of the Universe started.  The location of  Bo\"o I dSph is close enough to the Milky Way, however, it is possible to obtain the magnitudes and colours of MSTO stars with small photometric errors, $\Delta (V-Ic)_{0,MSTO}=0.03$.  Therefore, Bo\"o I dSph is the best target to seek for the possible age spread.

The colour width of MSTO are shown in Figures \ref{fig: CMDs of 4region of Boo1} and \ref{fig: bright MS histogram of 4region of Boo1}.  Figure \ref{fig: CMDs of 4region of Boo1} shows the CMDs of the four regions locating at the elliptical distance $r$ of (a) $0\arcmin<r<6\arcmin$, (b) $6\arcmin<r<10\arcmin$, (c) $10\arcmin <r<14\arcmin$, and (d) $14\arcmin<r<20\arcmin$.  Padova isochrone of Z=0.0001 and 13.7 Gyr is overlaid as a solid line.  These four CMDs look quite similar, and the number of stars belonging to Bo\"o I dSph decreases from the innermost (Figure \ref{fig: CMDs of 4region of Boo1}a) to the outside of half-light radius region (Figure \ref{fig: CMDs of 4region of Boo1}d). 

The observed width of MSTO is the convolution of the intrinsic width and photometric errors.  The intrinsic width of the MSTO could be broad due to multiple stellar populations \citep[e.g.,][]{2003AJ....126..218M,2004ApJ...605L.125B}, and the multiplicity of stellar populations can be examined from the width of the MSTO.  Figure \ref{fig: bright MS histogram of 4region of Boo1} shows the colour distributions of MSTO stars in the magnitude range of $22.7 < \vo < 22.9$, found in four regions shown in Figure \ref{fig: CMDs of 4region of Boo1}.  The solid lines are Gaussian distributions with $\sigma=0.03$, equal to the photometric error estimated from the artificial star tests.  The colour distributions of MSTO stars are all well represented by the photometric errors alone, which strongly suggests that Bo\"o I dSph has no intrinsic age spread, at least in the limit of Suprime-Cam photometry.  A Kolmogorov-Smirnov test is applied to confirm that the MSTO colour distribution of four regions are the same as the distribution produced by the photometric errors alone.  From this result, we conclude that Bo\"o I dSph has a single old stellar population. 


\begin{figure*}[th] 
 \begin{center}
 \includegraphics[width=400pt]{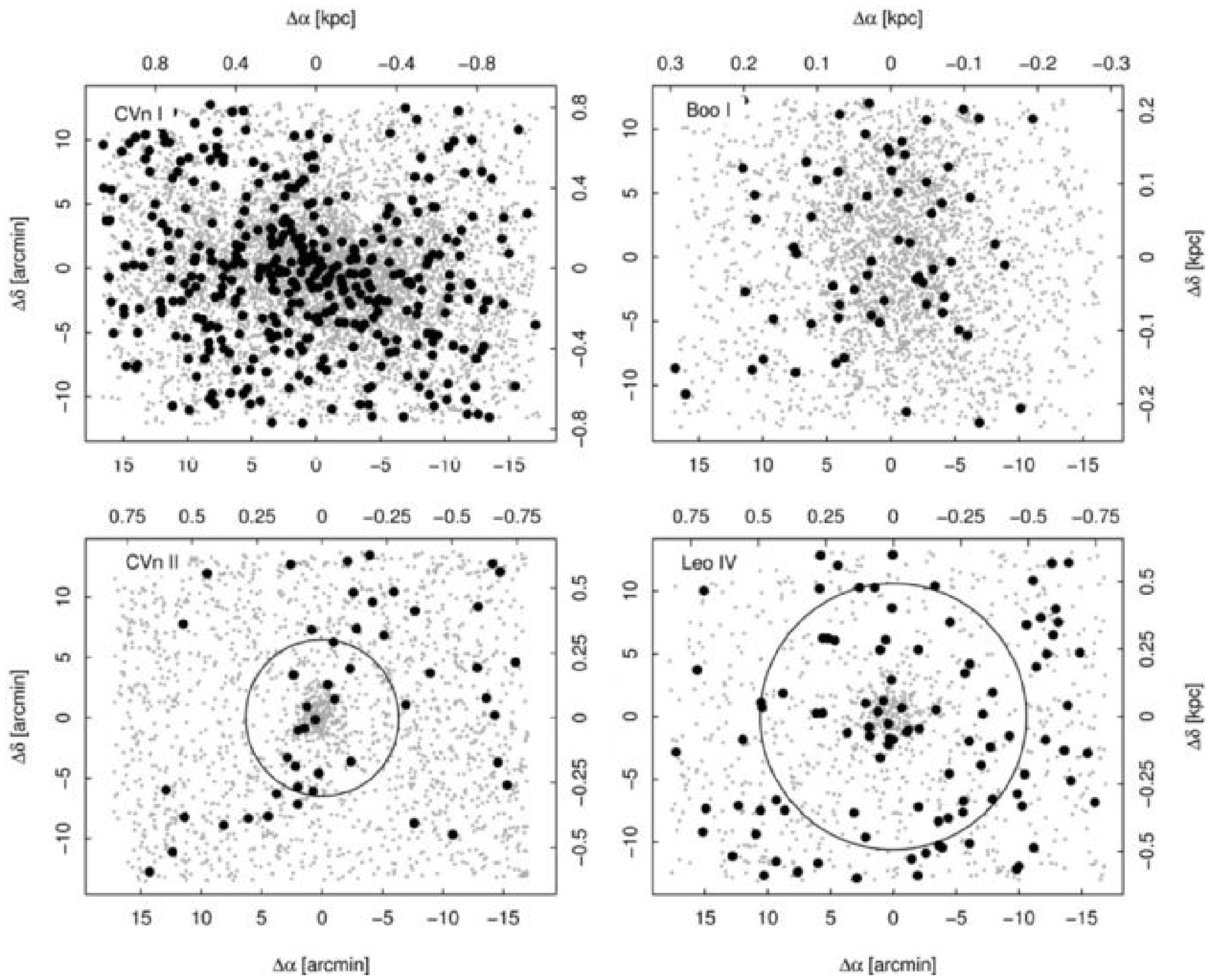}
 \caption{The spatial distributions of BS candidates found in the CVn I (upper-left), Bo\"o I (upper-right), CVn II (lower-left), and Leo IV dSphs (lower-right). The tidal radii of CVn II and Leo IV dSphs are shown as solid lines. }
  \label{fig: bs spatial}
  \end{center}
\end{figure*}


\subsection{Blue Stragglers in UFD galaxies}  \label{subsec: BS}

BS candidates are found in all four UFDs and distributed throughout the galaxies.  Figure \ref{fig: bs spatial} shows the spatial distribution of BS candidates of each UFD galaxy, which shows no sign of concentration toward the galaxy centre and clumpy distribution.  The BS criterion area in CMD which is shown as gray box above the MSTO in Figure \ref{fig: pop CMD of CVN1}b,  \ref{fig: pop CMD of BOO1}b, \ref{fig: pop CMD of CVN2}b, and \ref{fig: pop CMD of LEO4}b, is contaminated by unresolved background objects and foreground white dwarf stars.  These objects are seen outside of tidal radii of CVn II ($r > 0.29$~kpc) and Leo IV dSphs ($r > 0.48$~kpc) shown as the solid lines in Figure \ref{fig: bs spatial}.  

BS stars are universally found in Galactic globular clusters, open clusters, halo field, and in several Galactic dSphs \citep[e.g.,][]{1995ARA&A..33..133B, 2004ApJ...604L.109P, 2007MNRAS.380.1127M}.  They can form from stellar collisions in high density regions, or through mass transfer in isolated primordial binaries.  \citet{2006MNRAS.373..361M} suggested that the spatial distribution of BS stars in globular clusters provides strong hints to their origin.  The spatial distribution of collisional BS stars shows a peak at the cluster centre.  On the other hand, the mass-transfer BS stars follow the distribution of primordial binaries whose spatial distribution is similar to those of HB and RGB.
In case of dSphs galaxies with complex stellar populations, these blue stars can be either genuine BS stars or ordinary young ($\sim$ several Gyr) MS stars.  However, young populations are expected to be distributed more clumpy and centrally concentrated.  Therefore, the distribution of BS candidates in UFDs implies that these stars are not young MS stars, but mass-transfer BS stars which evolved from primordial binaries, as found in other dSphs \citep[e.g.][]{2007MNRAS.380.1127M, 2008A&A...487..103O}.

\subsection{Comparison with the Classical dSphs}

The stellar populations of UFDs are, in principle, quite old and metal-poor and are similar to those of the old metal-poor Galactic globular clusters.  Bo\"o I dSph shows no intrinsic spread in the width of MSTO, which indicates that Bo\"o I dSph experienced very short period of star formation.  CVn II and Leo IV dSphs, which are fainter than Bo\"o I dSph, are as old as Bo\"o I dSph.  The same is true for the other faint UFDs, Ursa Major I, II, Coma Berenices, and Hercules dSphs \citep{2008A&A...487..103O, 2008ApJ...672L..13M, 2009ApJ...704..898S, 2010AJ....140..138M}.  
The brightest UFD, CVn I dSph shows a slightly younger age compared with the other UFDs and different spatial distribution of BHB and RHB stars., implying the population gradient in this galaxy.  CVn I dSph shows also two kinematically distinct populations \citep{2006MNRAS.373L..70I, 2007MNRAS.380..281M}.  This population complexity is similar to those found in the brighter classical dSphs.  In general, the brighter Galactic satellites have more complex stellar populations than that of the fainter Galactic satellites.  

Two factors could be considered to make this population difference. The first one is accretion times and the second is potential depths of satellites.  
If fainter satellites had been captured by the Milky Way before they finished star formation,  the remains of the gas in the satellites were likely to be removed by the ram-pressure stripping at the accretion.  While they could hardly continue star formation after that, other small galaxies kept forming stars and became bright until they lost gas.  Furthermore, early accreted satellites had generally experienced a more significant mass loss than those accreted later, so that they became faint \citep{2004MNRAS.355..819G}.  
This can explain why UFDs are composed of only a few old stars, and why brighter dSphs show complex stellar populations.  
Numerical simulations, however, expect that the locations of early infalling satellites are likely to be closer to the Galactic centre than those of later infalling satellites \citep{2010MNRAS.406..744C}.  In this  scenario, early infalling satellites are expected to be fainter and closer to the Milky Way as compared with later infalling and brighter satellites.  The real UFDs ($M_{V}<-6$) are widely distributed around the Milky Way (30-160kpc), and there is no obvious correlation between the stellar population complexities and the current distances of Galactic satellites.  Therefore, whether the early accretion caused progenitors to become present UFDs is somewhat controversial. 

If we assume that the brighter satellites belonged to the more massive DM halos than those of fainter ones at their initial star formation, the deeper potential made it possible to keep the gas and form the stars for the longer duration, against the suppression effects such as the ram-pressure,  tidal stripping, SNe feedback, and photo-evaporation by reionization.  Therefore, the satellite progenitors in higher-mass halos tended to become brighter dSphs which had more complex stellar populations than those in lower-mass halos.  
After accretion, the satellites in lower mass halos should also suffer from stronger tidal effects of the Milky Way, which could explain the current elongated and distorted shapes of UFDs.  The mass estimations of Local Group dSphs, based on the velocity dispersion profiles of galaxies, revealed that the masses within $r_{h}$ of dSphs are proportional to the sizes and luminosities \citep{2009ApJ...704.1274W}.  This result supports the idea that the population complexity of Galactic satellites are mainly due to the difference in the potential depths of progenitors, and the faintness of UFDs directly reflect their shallow potentials.  In this case, star formation in faint UFDs, especially at large distances, had not been regulated by Galactic tides, but had been regulated by either reionization or SNe feedback, or influences of other small galaxies.

\section{Summary}

From the deep and wide images taken with Subaru/ Suprime-Cam, we demonstrate the single old stellar population of faint UFDs, Bo\"o I, CVn II, and Leo IV dSphs as well as the population complexity of the relatively bright UFD, CVn I dSph.  We confirm that Bo\"o I dSph has no intrinsic colour spread in the width of MSTO, and no spatial difference in the CMD morphology.  CVn I dSph, on the other hand, shows the relatively younger age ($\sim$ 12.6 Gyr), and different spatial distributions of BHB and RHB stars, implying the population gradient.  The spatial distributions of BS candidates in UFDs reveal that they are not young MS stars but mass-transfer BS stars.  These results indicate that the gases in the UFD progenitors were removed more effectively than the those of brighter dSphs when the initial star formation occurred.  This is reasonable if the progenitors of UFDs belong to less massive halos than those of brighter dSphs at that moment. 

The wide range of tidal radii and the distorted shapes of UFDs also imply that UFDs are strongly affected from Galactic tides.  We covered the region beyond $r_{t}$ of CVn II and Leo IV dSphs.  Although there are no extra stellar streams nor tidal debris around the galaxies, the radial profiles shows stellar overdensities at the edge of these galaxies.  Bo\"o I and CVn I dSphs show elongated morphologies, however, the observed areas are not enough to reveal the real extent.  Further wide and deep observations are required to clarify whether the highly elongated UFDs have spheroidal shapes or have shapes like stellar streams.  Our recent observation of the Hercules dSph, which covered the outer region of the galaxy, will provide an answer. 

We demonstrate that UFDs are composed of old stellar populations.  However it is still unclear which mechanism make such a faint galaxy.   Explorations of further faint nearby UFDs and isolated distant UFDs are crucial to reveal the origin.  Leo T dwarf is the only instance of the isolated UFD, so far, and is one of the brightest UFDs.  If there are numerous undiscovered UFDs in Local Group, the stellar ages in these galaxies are clue to the understanding the regulation mechanisms of star formations in small galaxies at the reionization epoch.

\acknowledgments
The authors are grateful to the observatory staff of the Subaru Telescope.  We wish to express our gratitude to the anonymous referee for very helpful suggestions and comments.  S.O. special thanks to M. Iye for great support and comment.  This work is supported by a Grant-in-Aid for Science Research (No.19540245 and No.21-8816) by the Japanese Ministry of Education, Culture, Sports, Science and Technology.

\clearpage


\begin{thebibliography}{}
\bibitem[\protect\citeauthoryear{Bailyn}{1995}]{1995ARA&A..33..133B} Bailyn C.~D., 1995, ARA\&A, 33, 133 
\bibitem[\protect\citeauthoryear{Battaglia et al.}{2006}]{2006A&A...459..423B} Battaglia G., et al., 2006, A\&A, 459, 423
\bibitem[\protect\citeauthoryear{Battaglia et al.}{2008}]{2008ApJ...681L..13B} Battaglia G., Helmi A., Tolstoy E., Irwin M., Hill V., Jablonka P., 2008, ApJ, 681, L13
\bibitem[\protect\citeauthoryear{Bedin et al.}{2004}]{2004ApJ...605L.125B} Bedin L.~R., Piotto G., Anderson J., Cassisi S., King I.~R., Momany Y., Carraro G., 2004, ApJ, 605, L125
\bibitem[\protect\citeauthoryear{Bellazzini et al.}{2004}]{2004A&A...424..199B} Bellazzini M., Ferraro F.~R., Sollima A., Pancino E., Origlia L., 2004, A\&A, 424, 199 
\bibitem[\protect\citeauthoryear{Belokurov et al.}{2006}]{2006ApJ...647L.111B} Belokurov V., et al., 2006, ApJ, 647, L111
\bibitem[\protect\citeauthoryear{Belokurov et al.}{2007}]{2007ApJ...654..897B} Belokurov V., et al., 2007b, ApJ, 654, 897
\bibitem[\protect\citeauthoryear{Bertin \& Arnouts}{1996}]{1996A&AS..117..393B} Bertin E., Arnouts S., 1996, A\&AS, 117, 393 
\bibitem[\protect\citeauthoryear{Bertin}{2006}]{2006ASPC..351..112B} Bertin E., 2006, ASPC, 351, 112
\bibitem[\protect\citeauthoryear{Cacciari \& Clementini}{2003}]{2003LNP...635..105C} Cacciari C., Clementini G., 2003, Stellar Candles for the Extragalactic Distance Scale (Lecture Notes in Physics 635), ed. D. Alloin \& W. Gieren (Berlin: Springer), 105
\bibitem[\protect\citeauthoryear{Cardelli, Clayton \& Mathis}{1989}]{1989ApJ...345..245C} Cardelli J.~A., Clayton G.~C., Mathis J.~S., 1989, ApJ, 345, 245
\bibitem[\protect\citeauthoryear{Carrera et al.}{2002}]{2002AJ....123.3199C} Carrera R., Aparicio A., Mart{\'{\i}}nez-Delgado D., Alonso-Garc{\'{\i}}a J., 2002, AJ, 123, 3199
\bibitem[\protect\citeauthoryear{Clem, Vanden Berg \& Stetson}{2008}]{2008AJ....135..682C} Clem J.~L., Vanden Berg D.~A., Stetson P.~B., 2008, AJ, 135, 682
\bibitem[\protect\citeauthoryear{Coleman et al.}{2007}]{2007ApJ...668L..43C} Coleman M.~G., et al., 2007, ApJ, 668, L43 
\bibitem[\protect\citeauthoryear{Cooper et al.}{2010}]{2010MNRAS.406..744C} Cooper A.~P., et al., 2010, MNRAS, 406, 744
\bibitem[\protect\citeauthoryear{Gao et al.}{2004}]{2004MNRAS.355..819G} Gao L., White S.~D.~M., Jenkins A., Stoehr F., Springel V., 2004, MNRAS, 355, 819 
\bibitem[\protect\citeauthoryear{Gilmore et al.}{2007}]{2007ApJ...663..948G} Gilmore, G., Wilkinson, M.~I., Wyse, R.~F.~G., Kleyna, J.~T., Koch, A., Evans, N.~W., \& Grebel, E.~K.\ 2007, \apj, 663, 948 
\bibitem[\protect\citeauthoryear{Girardi et al.}{2005}]{2005A&A...436..895G} Girardi L., Groenewegen M.~A.~T., Hatziminaoglou E., da Costa L., 2005, A\&A, 436, 895 
\bibitem[\protect\citeauthoryear{Greco et al.}{2008}]{2008ApJ...675L..73G} Greco C., et al., 2008, ApJ, 675, L73
\bibitem[\protect\citeauthoryear{Harris}{1996}]{1996AJ....112.1487H} Harris W.~E., 1996, AJ, 112, 1487
\bibitem[\protect\citeauthoryear{Ibata et al.}{2006}]{2006MNRAS.373L..70I} Ibata R., Chapman S., Irwin M., Lewis G., Martin N., 2006, MNRAS, 373, L70 
\bibitem[\protect\citeauthoryear{Johnson \& Bolte}{1998}]{1998AJ....115..693J} Johnson J.~A., Bolte M., 1998, AJ, 115, 693
\bibitem[\protect\citeauthoryear{Jordi, Grebel \& Ammon}{2006}]{2006A&A...460..339J} Jordi K., Grebel E.~K., Ammon K., 2006, A\&A, 460, 339 
\bibitem[\protect\citeauthoryear{King}{1962}]{1962AJ.....67..471K} King I., 1962, AJ, 67, 471
\bibitem[\protect\citeauthoryear{Kirby et al.}{2008}]{2008ApJ...685L..43K} Kirby E.~N., Simon J.~D., Geha M., Guhathakurta P., Frebel A., 2008, ApJ, 685, L43
\bibitem[\protect\citeauthoryear{Kuehn et al.}{2008}]{2008ApJ...674L..81K} Kuehn C., et al., 2008, ApJ, 674, L81
\bibitem[\protect\citeauthoryear{Landolt}{1992}]{1992AJ....104..340L} Landolt A.~U., 1992, AJ, 104, 340
\bibitem[\protect\citeauthoryear{Lee, Freedman \& Madore}{1993}]{1993ApJ...417..553L} Lee M.~G., Freedman W.~L., Madore B.~F., 1993, ApJ, 417, 553
\bibitem[\protect\citeauthoryear{Mapelli et al.}{2006}]{2006MNRAS.373..361M} Mapelli M., Sigurdsson S., Ferraro F.~R., Colpi M., Possenti A., Lanzoni B., 2006, MNRAS, 373, 361
\bibitem[\protect\citeauthoryear{Mapelli et al.}{2007}]{2007MNRAS.380.1127M} Mapelli M., Ripamonti E., Tolstoy E., Sigurdsson S., Irwin M.~J., Battaglia G., 2007, MNRAS, 380, 1127 
\bibitem[\protect\citeauthoryear{Marigo et al.}{2008}]{2008A&A...482..883M} Marigo P., Girardi L., Bressan A., Groenewegen M.~A.~T., Silva L., Granato G.~L., 2008, A\&A, 482, 883
\bibitem[\protect\citeauthoryear{Martin et al.}{2007}]{2007MNRAS.380..281M} Martin N.~F., Ibata R.~A., Chapman S.~C., Irwin M., Lewis G.~F., 2007, MNRAS, 380, 281
\bibitem[\protect\citeauthoryear{Martin et al.}{2008a}]{2008ApJ...672L..13M} Martin N.~F., et al., 2008a, ApJ, 672, L13
\bibitem[\protect\citeauthoryear{Martin, de Jong \& Rix}{2008b}]{2008ApJ...684.1075M} Martin N.~F., de Jong J.~T.~A., Rix H.-W., 2008b, ApJ, 684, 1075
\bibitem[\protect\citeauthoryear{Mateo}{1998}]{1998ARA&A..36..435M} Mateo M.~L., 1998, ARA\&A, 36, 435 
\bibitem[\protect\citeauthoryear{Monelli et al.}{2003}]{2003AJ....126..218M} Monelli M., et al., 2003, AJ, 126, 218
\bibitem[\protect\citeauthoryear{Moretti et al.}{2009}]{2009ApJ...699L.125M} Moretti M.~I., et al., 2009, ApJ, 699, L125
\bibitem[\protect\citeauthoryear{Mu{\~n}oz et al.}{2010}]{2010AJ....140..138M} Mu{\~n}oz, R.~R., Geha, M., \& Willman, B.\ 2010, \aj, 140, 138 
\bibitem[\protect\citeauthoryear{Mayer et al.}{2002}]{2002MNRAS.336..119M} Mayer L., Moore B., Quinn T., Governato F., Stadel J., 2002, MNRAS, 336, 119 
\bibitem[\protect\citeauthoryear{Norris et al.}{2008}]{2008ApJ...689L.113N} Norris J.~E., Gilmore G., Wyse R.~F.~G., Wilkinson M.~I., Belokurov V., Evans N.~W., Zucker D.~B., 2008, ApJ, 689, L113
\bibitem[\protect\citeauthoryear{Okamoto et al.}{2008}]{2008A&A...487..103O} Okamoto S., Arimoto N., Yamada Y., Onodera M., 2008, A\&A, 487, 103
\bibitem[\protect\citeauthoryear{Ouchi et al.}{2004}]{2004ApJ...611..660O} Ouchi M., et al., 2004, ApJ, 611, 660
\bibitem[\protect\citeauthoryear{Paust, Chaboyer, \& Sarajedini}{2007}]{2007AJ....133.2787P} Paust N.~E.~Q., Chaboyer B., Sarajedini A., 2007, AJ, 133, 2787 
\bibitem[\protect\citeauthoryear{Pe{\~n}arrubia et al.}{2010}]{2010MNRAS.406.1290P} Pe{\~n}arrubia J., Benson A.~J., Walker M.~G., Gilmore G., McConnachie A.~W., Mayer L., 2010, MNRAS, 406, 1290
\bibitem[\protect\citeauthoryear{Piotto et al.}{2004}]{2004ApJ...604L.109P} Piotto G., et al., 2004, ApJ, 604, L109 
\bibitem[\protect\citeauthoryear{Salaris, Chieffi \& Straniero}{1993}]{1993ApJ...414..580S} Salaris M., Chieffi A., Straniero O., 1993, ApJ, 414, 580
\bibitem[\protect\citeauthoryear{Sand et al.}{2009}]{2009ApJ...704..898S} Sand, D.~J., Olszewski, E.~W., Willman, B., et al.\ 2009, \apj, 704, 898 
\bibitem[\protect\citeauthoryear{Sand et al.}{2010}]{2010ApJ...718..530S} Sand, D.~J., Seth, A., Olszewski, E.~W., Willman, B., Zaritsky, D., \& Kallivayalil, N.\ 2010, \apj, 718, 530 
\bibitem[\protect\citeauthoryear{Schlegel, Finkbeiner \& Davis}{1998}]{1998ApJ...500..525S} Schlegel D.~J., Finkbeiner D.~P., Davis M., 1998, ApJ, 500, 525 
\bibitem[\protect\citeauthoryear{Sersic}{1968}]{1968adga.book.....S} Sersic J.~L., 1968, Cordoba, Argentina, ``{\it Atlas de galaxias australes}''
\bibitem[\protect\citeauthoryear{Shetrone et al.}{2003}]{2003AJ....125..684S} Shetrone M., Venn K.~A., Tolstoy E., Primas F., Hill V., Kaufer A., 2003, AJ, 125, 684
\bibitem[\protect\citeauthoryear{Stetson}{1987}]{1987PASP...99..191S} Stetson P.~B., 1987, PASP, 99, 191
\bibitem[\protect\citeauthoryear{Stobie}{1980}]{1980JBIS...33..323S} Stobie R.~S., 1980, J. British Interplanet. Soc., 33, 323
\bibitem[\protect\citeauthoryear{Susa \& Umemura}{2004}]{2004ApJ...600....1S} Susa H., Umemura M., 2004, ApJ, 600, 1 
\bibitem[\protect\citeauthoryear{Tolstoy et al.}{2004}]{2004ApJ...617L.119T} Tolstoy E., et al., 2004, ApJ, 617, L119
\bibitem[\protect\citeauthoryear{Tolstoy, Hill \& Tosi}{2009}]{2009ARA&A..47..371T} Tolstoy E., Hill V., Tosi M., 2009, ARA\&A, 47, 371
\bibitem[\protect\citeauthoryear{Walker et al.}{2009}]{2009ApJ...704.1274W} Walker M.~G., Mateo M., Olszewski E.~W., Pe{\~n}arrubia J., Wyn Evans N., Gilmore G., 2009, ApJ, 704, 1274 
\bibitem[\protect\citeauthoryear{Willman et al.}{2005}]{2005ApJ...626L..85W} Willman B., et al., 2005, ApJ, 626, L85 
\bibitem[\protect\citeauthoryear{Yagi et al.}{2002}]{2002AJ....123...66Y} Yagi M., Kashikawa N., Sekiguchi M., Doi M., Yasuda N., Shimasaku K., Okamura S., 2002, AJ, 123, 66
\bibitem[\protect\citeauthoryear{Zucker et al.}{2006}]{2006ApJ...643L.103Z} Zucker D.~B., et al., 2006, ApJ, 643, L103

\end{thebibliography}
\end{document}